\documentclass[draftclsnofoot,onecolumn,12pt]{IEEEtran}
\usepackage{graphicx}
\usepackage{epstopdf}

\usepackage{subfigure}
\usepackage{booktabs}
\usepackage{caption}
\usepackage{color}
\usepackage{bm}
\usepackage{CJK}
\usepackage{indentfirst}
\usepackage{amsmath}
\usepackage{amssymb}
\usepackage{float}
\usepackage{multirow}
\usepackage{algorithm}
\usepackage{algpseudocode}
\usepackage{amsmath}
\usepackage{flushend}
\usepackage{geometry}
\newtheorem{theorem}{Theorem}
\newtheorem{lemma}{Lemma}
\newtheorem{property}{Property}

\begin{document}
\title{Adaptive Bit Partitioning for Reconfigurable Intelligent Surface Assisted FDD Systems with Limited Feedback}
%% author names and affiliations
\author{\IEEEauthorblockN{Weicong~Chen, Chao-Kai Wen, Xiao Li, and~Shi~Jin}\\
\thanks{{Weicong Chen, Xiao Li, and Shi Jin are with the National Mobile Communications
Research Laboratory, Southeast University, Nanjing 210096, China (e-mail: cwc@seu.edu.cn; li\_xiao@seu.edu.cn; jinshi@seu.edu.cn).} }
\thanks{Chao-Kai Wen is with the Institute of Communications Engineering,
National Sun Yat-sen University, Kaohsiung City 80424, Taiwan (e-mail:
chaokai.wen@mail.nsysu.edu.tw).}
}

%% make the title area
\maketitle
\begin{abstract}
In frequency division duplexing systems, the base station (BS) acquires downlink channel state information (CSI) via channel feedback, which has not been adequately investigated in the presence of RIS. In this study, we examine the limited channel feedback scheme by proposing a novel cascaded codebook and an adaptive bit partitioning strategy. The RIS segments the channel between the BS and mobile station into two sub-channels, each with line-of-sight (LoS) and non-LoS (NLoS) paths. To quantize the path gains, the cascaded codebook is proposed to be synthesized by two sub-codebooks whose codeword is cascaded by LoS and NLoS components. This enables the proposed cascaded codebook to cater the different distributions of LoS and NLoS path gains by flexibly using different feedback bits to design the codeword structure. On the basis of the proposed cascaded codebook, we derive an upper bound on ergodic rate loss with maximum ratio transmission and show that the rate loss can be cut down by optimizing the feedback bit allocation during codebook generation. To minimize the upper bound, we propose a bit partitioning strategy that is adaptive to diverse environment and system parameters. Extensive simulations are presented to show the superiority and robustness of the cascaded codebook and the efficiency of the adaptive bit partitioning scheme.
\end{abstract}

%%%%%%%%%%%%%%%%%%%%%%%%%%%%%%%%%%%%%%%%%%%%%%%%%%%%%%%%%%%%%%%%%%%%%%%%%%%%%%%%%%%%%%%%%%%%%%
\section{Introduction}
The unprecedented popularity of wireless communicating devices inflicts enormous challenges on existing wireless communication systems. Future wireless networks are envisioned to be turned into a distributed intelligent communications, sensing, and computing platform \cite{smart-radio}. Among various advanced technologies that emerge to realize ubiquitous wireless connectivity, reconfigurable intelligent surface\footnote{In the literature, RIS is also termed as large intelligent surface (LIS) \cite{Y. Han-LIS} or intelligent reflecting surface (IRS) \cite{Q. Wu-IRS}.} (RIS) has recently received fervent attention in both academia and industry. The concept of RIS is extended from metasurface, which is composed of a large array of passive scattering elements with a specially designed physical structure \cite{S. Gong}. The introduction of RISs has the potential to simplify the traditional transceiver structure \cite{W. Tnag_1}--\cite{W. Tnag_4} and realize a programmable smart radio environment in a low-cost way \cite{smart-radio}, \cite{C. Liaskos}.\par

Recently, numerous works \cite{App_EE}--\cite{App-IoT} focused on the diverse RIS-assisted applications in wireless communication systems. In \cite{App_EE}, the authors showed that utilizing the proposed RIS-based
resource allocation methods can achieve $300\%$ higher energy efficiency in comparison with the use of regular multi-antenna amplify-and-forward relaying. The authors in \cite{App_scerecy} proposed the use of RIS to create friendly multipaths for directional modulation so that two confidential bit streams can be transmitted from Alice to Bob, which can significantly enhance the secrecy rate of directional modulation. To reliably transmit information, a reflecting modulation scheme was proposed in \cite{App-reflecting-modulation} for RIS-assisted communications where both reflecting patterns and transmit signals can carry information. For coverage enhancement in mmWave communication systems, RISs were introduced to alleviate the significant path loss and severe blockage in \cite{App-mmWave-1} and \cite{App-mmWave-2}, respectively. Moreover, \cite{App-D2D} and \cite{App-IoT} employed RISs into device-to-device and Internet-of-Things communication systems, respectively, to support super-massive access. The aforementioned works assumed that the perfect channel state information (CSI) is available, which, however, is difficult to realize in practical systems due to the large number of nearly passive elements that cannot perform signal processing.\par

To tackle new challenges on channel estimation in RIS-assisted systems, \cite{est-atomic} proposed a two-stage channel estimation scheme in mmWave band using atomic norm minimization to sequentially estimate the channel parameters. In THz communications powered by RIS, \cite{est-THz} designed a novel hierarchical search codebook and realized the channel estimation by low-complexity beam training. For RIS-assisted multi-user broadband systems employing OFDMA, in \cite{est-partial}, a channel estimation scheme that  exploits a key property that all users share partially identical channel was proposed to enhance training efficiency and support more users. By regarding the channels from the base station (BS) to the RIS and from the RIS to the mobile station (MS) as a cascaded channel, the channel estimation problem was formulated into a sparse channel matrix recovery problem in \cite{est-CS} using the compressed sensing technique. Besides algorithms, a novel RIS architecture based on sparse channel sensors was proposed in \cite{est-sensor} to enable channel estimation at the RIS.\par

Downlink CSI is well known to be essential for the BS to design transmission schemes. In frequency division duplex (FDD) systems, downlink CSI is estimated at the MS and then fed back to the BS. Although channel feedback has been widely studied in current wireless communication \cite{Nihar}--\cite{Ramta_2}, it cannot directly apply to the RIS-assisted counterpart because the channel matrix involving the RIS is extremely large, which results in prohibitive feedback overhead. For the first time, channel feedback was investigated in \cite{feedback-RIS} for RIS-assisted wireless communications. The channel path gains quantization and feedback in \cite{feedback-RIS} were performed with naive random vector quantization (RVQ) codebook, which requires the cascaded path gains to be independently identically distributed. This requirement, however, cannot be satisfied because cascaded path gains have disparate distribution when the line-of-sight (LoS) and non-LoS (NLoS) paths are considered. Theoretical performance analysis was not carried out as well.\par

Given the observation that LoS and NLoS path gains cannot be quantized in the same way, in this study, we propose a novel cascaded codebook and an adaptive bit partitioning strategy for channel feedback in RIS-assisted communication systems with limited feedback bits. Considering that the information of channel path directions varies more slowly than path gains \cite{Naive_RVQ} and is spatial reciprocal in the uplink and downlink \cite{Han_FDD}, the BS can easily obtain the path directions; thus, we focus on the path gains feedback. The contributions of our work are summarized as follows.
\begin{itemize}
  \item A novel cascaded codebook is proposed for path gain feedback in RIS-assisted wireless communication systems. The introduction of RIS segments the channel between the BS and MS into BS--RIS and RIS--MS sub-channels and each segmented channel has LoS and NLoS paths. Therefore, we synthesize the cascaded codebook by two sub-codebooks whose codewords are cascaded by LoS and NLoS components. This approach enables the cascaded codebook to be used in a variety of scenarios thanks to its alterable structure that captures the different distribution features of path gains.
  \item On the basis of the proposed cascaded codebook and the optimal reflection coefficients design for RIS, a theoretical upper bound on ergodic rate loss is derived with maximum ratio transmission (MRT) when feedback bits are limited. The closed-form upper bound describes the path gain quantization error incurred by limited feedback and is the function of separated feedback bits. Hence, it shows the potential of reducing ergodic rate loss by optimizing the feedback bit assignment.
  \item A bit partitioning strategy that divides feedback bits into four parts to generate sub-codebooks is proposed. The upper bound on ergodic rate loss can be minimized by proper bit allocation. Through the closed-form expressions for bit partitioning, we show that the number of RIS elements, Rician factors, and number of paths jointly determine the allocation criteria. Rician factors and number of paths play critical roles in bit allocation as they describe the distribution characteristic of paths. Meanwhile, the effect of the number of RIS elements on the allocation criteria is embodied in the difference of averaged signal powers received from the cascaded LoS path and other paths.
  \item Numerical results are carried out to show the superiority of the proposed cascaded codebook and adaptive bit partitioning strategy. It is demonstrated that the cascaded codebook outperforms the naive RVQ codebook with more robust ergodic rate and lower feedback bits requirement. The bit partitioning strategy that adapts to different environment and system parameters is presented. With the use of our bit allocation scheme, the ergodic rate loss is verified to be efficiently reduced.
\end{itemize}\par
The rest of this paper is organized as follows. The system model is described in Section \ref{sec:2}. The cascaded codebook for limited feedback and analysis for ergodic rate loss are presented in Section \ref{sec:4}. The bit partitioning strategy is given in Section \ref{sec:5}, and numerical results are used to verify our proposal in Section \ref{sec:6}. Finally, conclusions are drawn in Section \ref{sec:7}.\par

\emph{Notations}: The lowercase and uppercase of a letter represent the vector and matrix, respectively. $\lceil\cdot\rceil$ and $\lfloor \cdot \rfloor$ are the integer ceiling and floor, respectively. The superscripts $(\cdot)^T$, $(\cdot)^*$, and $(\cdot)^H$ denote the transpose, conjugate, and conjugated-transpose operation, respectively. The Kronecker and element-wise product are denoted respectively by $\otimes$ and $\odot$. ${\mathbb E}\{\cdot\}$ is to calculate the mean. $|\cdot|$ and $\|\cdot\|$ are used to indicate the absolute value and Euclidean norm, respectively. ${\rm diag}(a_1,a_2,\cdots,a_N)$ indicates a diagonal matrix with diagonal elements $a_i$, $i=1,\cdots,N$. ${\mathcal{U}}\left[ {a,b } \right]$ represents the uniform distribution between $a$ and $b$.

%%%%%%%%%%%%%%%%%%%%%%%%%%%%%%%%%%%%%%%%%%%%%%%%%%%%%%%%%%%%%%%%%%%%%%%%%%%%%%%%%%%%%%%%%%%%%%
\section{System Model}\label{sec:2}
\begin{figure}[!t]
\centering
    \includegraphics[width=0.5\textwidth]{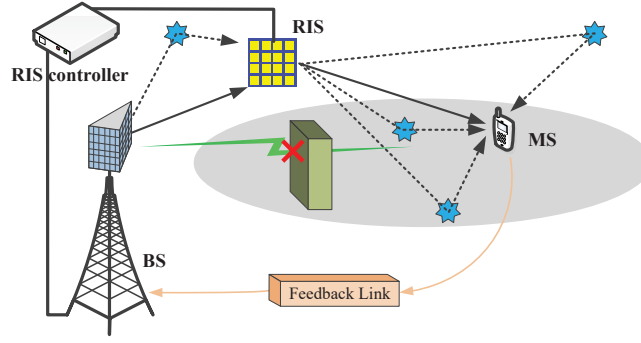}
\caption{RIS-assisted FDD system where downlink CSI is fed back to the BS via feedback link.}
\label{Fig.system_model}
\end{figure}

In this section, we describe the RIS-assisted FDD system shown in Fig. \ref{Fig.system_model}, where the BS is assisted by an RIS to serve a single-antenna MS. Considering the blockage, we assume that there is no direct link between the BS and the MS, and thus the MS can only receive data from the BS through an RIS. In FDD systems, the downlink CSI is estimated at the MS and then fed back to the BS via feedback link. The RIS is controlled by the BS through an RIS controller to manipulate the electromagnetic response of incident waves. The BS (or the RIS) is equipped with a uniform planar array (UPA) with $N_{\rm B}=N_{\rm B,v}\times N_{\rm B,h}$ (or $N_{\rm R}=N_{\rm R,v}\times N_{\rm R,h}$) antennas (or RIS elements), where the subscripts $\rm v$ and $\rm h$ denote the vertical and horizontal system parameters, respectively. The vertical and horizontal antenna (or RIS element) spacings are denoted by $\Delta_{\rm B,v}$ and $\Delta_{\rm B,h}$ (or $\Delta_{\rm R,v}$ and $\Delta_{\rm R,h}$), respectively.

\par

\subsection{Channel Model}\label{sec:2.1}
In the typical deployment of RIS, the LoS link exists to improve the system performance. Therefore, we adopt the geometric model incorporating the Rician factor to describe the BS--RIS channel
\begin{equation}\label{eq:H}
  {\bf{H}} = \sqrt {\frac{{{K_{\rm{B}}}}}{{1 + {K_{\rm{B}}}}}} {{\bf{H}}_{\rm{L}}} + \sqrt {\frac{1}{{1 + {K_{\rm{B}}}}}} {{\bf{H}}_{\rm{N}}},
\end{equation}
where the Rician factor ${K_{\rm{B}}}$ denotes the power ratio of the LoS channel component ${{\bf{H}}_{\rm{L}}}$ and the NLoS component ${{\bf{H}}_{\rm{N}}}$. The LoS channel ${{\bf{H}}_{\rm{L}}}$ can be expressed as
\begin{equation}\label{eq:H_L}
  {{\bf{H}}_{\rm{L}}} = \sqrt {{N_{\rm{B}}}{N_{\rm{R}}}} {g_{{\rm{B}},{\rm{L}}}}{{\bf{a}}_{\rm{R}}}\left( {{\Theta _{{\rm{1,r}}}},{\Phi _{{\rm{1,r}}}}} \right){\bf{a}}_{\rm{B}}^H\left( {{\Theta _1},{\Phi _1}} \right),
\end{equation}
where ${g_{{\rm{B}},{\rm{L}}}} = {e^{j{\eta _{\rm{B}}}}}$ is the normalized complex gain of the LoS path with ${\eta _{\rm{B}}} \sim {\mathcal U}\left[ {0,2\pi } \right]$; and ${\Theta _{1{\rm{,r}}}} = \frac{{2\pi {\Delta _{\rm{R,v}}}}}{\lambda }\cos {\theta _{1{\rm{,r}}}}$, ${\Phi _{1{\rm{,r}}}} = \frac{{2\pi {\Delta _{\rm{R,h}}}}}{\lambda }\sin {\theta _{1{\rm{,r}}}}\sin {\phi _{1{\rm{,r}}}}$, ${\Theta _{1}} = \frac{{2\pi {\Delta _{\rm{B,v}}}}}{\lambda }\cos {\theta _{1}}$, and ${\Phi _{1}} = \frac{{2\pi {\Delta _{\rm{B,h}}}}}{\lambda }\sin {\theta _{1}}\sin {\phi _{1}}$ are spatial frequencies of the LoS path, where $(\theta_{1,{\rm r}}, \phi_{1,{\rm r}})$ and $(\theta_1, \phi_1)$ are the angle of arrivals (AoAs) of the RIS and the angle of departures (AoDs) of the BS, respectively. For convenience of notation, we use ${\bf a}_{X}(\Theta,\Phi) = {\bf a}_{X,{\rm v}}(\Theta)\otimes{\bf a}_{X,{\rm h}}(\Phi)$ to denote the array response of UPA, where
\begin{equation}\label{ea:ULA}
  {\bf a}_{X,y}(Z) =\frac{1}{{\sqrt {N_{X,y}} }}{\left[ {\begin{array}{*{20}{c}}
1&{{e^{jZ}}}& \cdots &{{e^{j\left( {N_{X,y} - 1} \right)Z}}}
\end{array}} \right]^T}
\end{equation}
for ${ X}\in \{{\rm R,B}\}$ and $(y,Z)\in\{({\rm v},\Theta),({\rm h},\Phi)\}$.
In \eqref{eq:H}, the NLoS channel can be expressed as ${{\bf{H}}_{\rm{N}}} = \sum\nolimits_{l = 2}^{{L_{\rm{B}}}} {{{\bf{H}}_{l,{\rm{N}}}}} $, where $L_{\rm B}$ is the number of paths in the BS--RIS channel and the $l$-th NLoS path can be expressed as
\begin{equation}\label{eq:H_l,N}
  {{\bf{H}}_{l,{\rm{N}}}} = \sqrt {\frac{{{N_{\rm{B}}}{N_{\rm{R}}}}}{{{L_{\rm{B}}} - 1}}} {g_{{\rm{B}},l}}{{\bf{a}}_{\rm{R}}}\left( {{\Theta _{l{\rm{,r}}}},{\Phi _{l{\rm{,r}}}}} \right){\bf{a}}_{\rm{B}}^H\left( {{\Theta _l},{\Phi _l}} \right),
\end{equation}
with ${g_{{\rm{B}},l}} \sim {\mathcal {CN}}\left( {0,1} \right)$ for $l\in \{2,\cdots,L_{\rm B}\}$ being the $l$-th path gain. For clarity of presentation, we merge \eqref{eq:H_L} and \eqref{eq:H_l,N} into \eqref{eq:H} to rewrite $\bf H$ as
\begin{equation}\label{eq:H_unit}
  {\bf{H}} = \sqrt {\frac{{{N_{\rm{B}}}{N_{\rm{R}}}}}{{\left( {1 + {K_{\rm{B}}}} \right)\left( {{L_{\rm{B}}} - 1} \right)}}} \sum\limits_{l = 1}^{{L_{\rm{B}}}} {{g_{{\rm{B}},l}}{{\bf{a}}_{\rm{R}}}\left( {{\Theta _{l{\rm{,r}}}},{\Phi _{l{\rm{,r}}}}} \right){\bf{a}}_{\rm{B}}^H\left( {{\Theta _{l}},{\Phi _l}} \right)} ,
\end{equation}
where ${g_{{\rm{B}},1}} = \sqrt {{K_{\rm{B}}}\left( {{L_{\rm{B}}} - 1} \right)} {g_{{\rm{B}},{\rm{L}}}}$.\par
Similarly, the RIS--MS channel can be expressed by
\begin{equation}\label{eq:h}
  {\bf{h}} = \sqrt {\frac{{{N_{\rm{R}}}}}{{\left( {1 + {K_{\rm{M}}}} \right)\left( {{L_{\rm{M}}} - 1} \right)}}} \sum\limits_{i = 1}^{{L_{\rm{M}}}} {{g_{{\rm{M}},i}}{{\bf{a}}_{\rm{R}}}\left( {{\Theta _{i{\rm{,t}}}},{\Phi _{i{\rm{,t}}}}} \right)},
\end{equation}
where $K_{\rm M}$ and $L_{\rm M}$ denote the Rician factor and number of paths in the RIS--MS channel, respectively; ${g_{{\rm{M}},1}} = \sqrt {{K_{\rm{M}}}\left( {{L_{\rm{M}}} - 1} \right)} {g_{{\rm{M,L}}}}$ with ${g_{{\rm{M,L}}}} = {e^{j{\eta _{\rm{M}}}}}$ being the normalized LoS path gain and ${\eta _{\rm{M}}} \sim {\mathcal U}\left[ {0,2\pi } \right]$; ${g_{{\rm{M}},i}} \sim {\mathcal {CN}}\left( {0,1} \right)$ for $i \in \left\{ {2,\cdots,{L_{\rm{M}}}} \right\}$ is the NLoS path gain; and ${\Theta _{i{\rm{,t}}}}=\frac{{2\pi {\Delta _{\rm{R,v}}}}}{\lambda }\cos {\theta _{i{\rm{,t}}}}$ and ${\Phi _{i{\rm{,t}}}}=\frac{{2\pi {\Delta _{\rm{R,v}}}}}{\lambda }\sin {\theta _{i{\rm{,t}}}}\sin{\phi_{i,{\rm t}}}$ are spatial frequencies of the $i$-th path, respectively, where $\theta_{i,{\rm t}}$ and $\phi_{i,{\rm t}}$
are the AoDs of RIS.
\subsection{Ergodic Rate}\label{sec:2.2}
In the RIS-assisted downlink transmission, the received signal of the MS is given by
\begin{equation}\label{eq:y}
  y = {{\bf{h}}^H}{\bf{\Psi Hf}}x + n,
\end{equation}
where ${\bf{\Psi }} = {\rm{diag}}\left( {\bm{\psi }} \right){\rm{ = diag}}( {{e^{j{\psi _1}}}, \cdots ,{e^{j{\psi _{{N_{\rm{R}}}}}}}} )$ is the reflection coefficient matrix of the RIS\footnote{Although the reflection coefficient of the RIS is shown to be depended on the incident angle of EM waves in \cite{AngleReci}, this characteristic has no impact on this paper since we design the reflection coefficient only for LoS path. The focus of our paper is the channel feedback, thus we neglect the angle-dependent property of RIS for sake of concise description.}, $\bf f$ denotes the precoding vector adopted by the BS, $x$ is the transmitted signal satisfying ${\mathbb E}\{ {{{| x |}^2}} \} = E$ with $E$ being the transmit power, and $n\sim {\mathcal {CN}}(0,\sigma^2)$ is the complex Gaussian noise with noise power $\sigma^2$.\par
We denote the cascaded BS--RIS--MS channel as ${\bf{h}}_{{\rm{eff}}}^H \buildrel \Delta \over = {{\bf{h}}^H}{\bf{\Psi H}}$ and rewrite it as
\begin{equation}\label{eq:h_eff}
  {\bf{h}}_{{\rm{eff}}}^H = {Q_1}{\left( {{{\bf{g}}_{\rm{B}}} \otimes {{\bf{g}}_{\rm{M}}}} \right)^H}{{\bf{A}}^H},
\end{equation}
where ${{\bf{g}}_{\rm{B}}} = {\left[ {{g_{{\rm{B}},1}}, \cdots ,{g_{{\rm{B}},{L_{\rm{B}}}}}} \right]^T}$ and ${{\bf{g}}_{\rm{M}}} = {\left[ {{g^*_{{\rm{M}},1}}, \cdots ,{g^*_{{\rm{M}},{L_{\rm{M}}}}}} \right]^T}$ denote the path gain vector for the BS--RIS and RIS--MS channel, respectively,
\begin{equation}\label{eq:Q_1}
  {Q_1} = \sqrt {\frac{{{N_{\rm{B}}}N_{\rm{R}}^2}}{{\left( {1 + {K_{\rm{M}}}} \right)\left( {1 + {K_{\rm{B}}}} \right)\left( {{L_{\rm{M}}} - 1} \right)\left( {{L_{\rm{B}}} - 1} \right)}}} ,
\end{equation}
and ${\bf{A}} = \left[ {{{\bf{a}}_{{\rm{E}},1}}, \cdots ,{{\bf{a}}_{{\rm{E}},{L_{\rm{M}}}{L_{\rm{B}}}}}} \right]$ is the cascaded path direction matrix whose $n$-th column is given by
\begin{equation}\label{eq:a_E,n}
  {{\bf{a}}_{{\rm{E}},n}} = {{\bf{a}}_{\rm{B}}}\left( {{\Theta _l},{\Phi _l}} \right){\bf{a}}_{\rm{R}}^H\left( {{\Theta _{l{\rm{,r}}}},{\Phi _{l{\rm{,r}}}}} \right){\rm{diag}}\left( {{{\bf{a}}_{\rm{R}}}\left( {{\Theta _{i{\rm{,t}}}},{\Phi _{i{\rm{,t}}}}} \right)} \right){{\bm{\psi }}^*}.
\end{equation}
Defining two functions
\begin{equation}\label{eq:two_func}
  \left\{ \begin{array}{l}
{v_{{L_{\rm{M}}}}}\left( n \right) \buildrel \Delta \over = \left\lceil {n/{L_{\rm{M}}}} \right\rceil \\
{h_{{L_{\rm{M}}}}}\left( n \right) \buildrel \Delta \over = n - \left( {\left\lceil {n/{L_{\rm{M}}}} \right\rceil  - 1} \right){L_{\rm{M}}}
\end{array} \right.,
\end{equation}
and then the relationships between subscripts $n$, $l$, and $i$ in \eqref{eq:a_E,n} are
\begin{equation}
  \left\{ \begin{array}{l}
n = \left( {l - 1} \right){L_{\rm{M}}} + i \\
l = {v_{{L_{\rm{M}}}}}\left( n \right)\\
i = {h_{{L_{\rm{M}}}}}\left( n \right)
\end{array} \right.,
\end{equation}\par
When Jensen's inequality is applied, the downlink ergodic rate is upper bounded as
\begin{equation}\label{eq:Jensen}
  {R} = {\mathbb E}\left\{ {{{\log }_2}\left( {1 + {{{\left| {{\bf{h}}_{{\rm{eff}}}^H{\bf{f}}} \right|}^2}}/{{{\sigma ^2}}}} \right)} \right\} \le {\log _2}\left( {1 + {{\mathbb E}{\left\{ {{{\left| {{\bf{h}}_{{\rm{eff}}}^H{\bf{f}}} \right|}^2}} \right\}}}/{{{\sigma ^2}}}} \right).
\end{equation}
In this syudy, we consider that the BS adopts MRT, that is, ${\bf{f}} = {{\bf{h}}_{{\rm{eff}}}}/\left\| {{{\bf{h}}_{{\rm{eff}}}}} \right\|$. The upper bound depends on the $\bf f$, which is designed using perfect CSI. In the FDD system, the downlink CSI is acquired through channel feedback by the MS, as shown in Fig. \ref{Fig.system_model}. Considering the limited feedback ability of the MS, only quantized CSI can be obtained by the BS. Hence, designing the MRT with quantized CSI will inevitably result in a rate loss. To reduce the rate loss, in the following section, we design a limited feedback scheme with a novel cascaded codebook and analyze its performance. An adaptive feedback bit partitioning algorithm based on a derived closed-form rate loss expression is proposed to further cut down the rate loss.

%%%%%%%%%%%%%%%%%%%%%%%%%%%%%%%%%%%%%%%%%%%%%%%%%%%%%%%%%%%%%%%%%%%%%%%%%%%%%%%%%%%%%%%%%%%%%%
\section{Cascade Codebook Design and Performance Analysis}\label{sec:4}
In this section, the closed-form upper bound on ergodic rate with perfect CSI is initially derived as baseline. We then design a cascaded codebook for limited CSI feedback. On the basis of the proposed codebook, the ergodic rate loss is analyzed and further upper bounded by considering the quantization error brought by limited feedback bits.\par
Denoting $P \buildrel \Delta \over = {{\mathbb E}{\{ {{{| {{\bf{h}}_{{\rm{eff}}}^H{\bf{f}}} |}^2}} \}}}$ for \eqref{eq:Jensen} and following the analysis procedure in \cite{SAoS}, we have
\begin{equation}\label{eq:P}
  P = \frac{{{N_{\rm{B}}}}}{{\left( {1 + {K_{\rm{M}}}} \right)\left( {1 + {K_{\rm{B}}}} \right)}}\left( {{K_{\rm{B}}}{K_{\rm{M}}}\sum\limits_{c = 1}^{{N_{\rm{R}}}} {\sum\limits_{s = 1}^{{N_{\rm{R}}}} {{e^{j\left( {{\Omega _{1,1,c}} - {\Omega _{1,1,s}}} \right)}}} }  + {N_{\rm{R}}}\left( {{K_{\rm{M}}} + {K_{\rm{B}}} + 1} \right)} \right).
\end{equation}
The general form of ${\Omega _{1,1,s}}$ in \eqref{eq:P} is given by
\begin{equation}\label{eq:OMIGA}
  {\Omega _{l,i,s}} = {\psi _s} + \left( {{v_{{N_{{\rm{R,h}}}}}}\left( s \right) - 1} \right)\left( {{\Theta _{l{\rm{,r}}}} - {\Theta _{i{\rm{,t}}}}} \right) + \left( {{h_{{N_{{\rm{R,h}}}}}}\left( s \right) - 1} \right)\left( {{\Phi _{l{\rm{,r}}}} - {\Phi _{i{\rm{,t}}}}} \right),
\end{equation}
where the subscripts $l$ and $i$ refer to parameters of the $l$-th path in the BS--RIS channel and the $i$-th path in the RIS--MS channel, respectively. On the basis of \eqref{eq:P} and \eqref{eq:OMIGA}, it is easy to conclude that when the MRT is adopted, the upper bound of the ergodic rate with perfect CSI can be maximized by making ${\Omega _{1,1,c}} - {\Omega _{1,1,s}} = 0$. Therefore, given the $c$-th RIS element to set its reflection phase $\psi _c^\# $ as a reference, the optimal refection phase of the $s$-th ($s\in \{1,\cdots,N_{\rm R}\}$) RIS element can be designed as
\begin{equation}\label{eq:optimal_phase}
  \psi _s^\#  = \psi _c^\#  + \left( {{v_{{N_{{\rm{R,h}}}}}}\left( c \right) - {v_{{N_{{\rm{R,h}}}}}}\left( s \right)} \right)\left( {{\Theta _{{\rm{1,r}}}} - {\Theta _{{\rm{1,t}}}}} \right) + \left( {{h_{{N_{{\rm{R,h}}}}}}\left( c \right) - {h_{{N_{{\rm{R,h}}}}}}\left( s \right)} \right)\left( {{\Phi _{{\rm{1,r}}}} - {\Phi _{{\rm{1,t}}}}} \right).
\end{equation}
Thus, the upper bound ergodic rate with perfect CSI is given by
\begin{equation}\label{eq:C_upper}
  {R_{{\rm{P,upper}}}} = {\log _2}\left( {1 + {{{P_{{\rm{opt}}}}}}/{{{\sigma ^2}}}} \right),
\end{equation}
where
\begin{equation}\label{eq:P-opt}
  {P_{{\rm{opt}}}} = \frac{{{N_{\rm{B}}}N_{\rm{R}}^2{K_{\rm{B}}}{K_{\rm{M}}} + {N_{\rm{B}}}{N_{\rm{R}}}\left( {{K_{\rm{M}}} + {K_{\rm{B}}} + 1} \right)}}{{\left( {1 + {K_{\rm{M}}}} \right)\left( {1 + {K_{\rm{B}}}} \right)}}.
\end{equation}\par
Designing the MRT at the BS requires downlink CSI, which is fed back by the MS after channel estimation. To feed back the CSI with limited bits, we should design a predetermined codebook so that an appropriate codeword can be selected to quantize the CSI as accurately as possible.

\subsection{Cascaded Codebook Design for Path Gain Feedback}\label{sec:3}
Introducing the RIS into wireless systems naturally segments the BS-MS channel into two sub-channels. When designing its precoding vector and reflection coefficient matrix for the RIS, it is essential for the BS to acquire the downlink CSI $\bf H$ and $\bf h$. Given the abundant elements integrated in the RIS, it is unaffordable for the MS to directly feed back the large dimensional $\bf H$ and $\bf h$ to the BS when the feedback ability is limited. Fortunately, parameterizing the channels as \eqref{eq:H_unit} and \eqref{eq:h} means that the BS can reconstruct the downlink channel as long as the MS feeds back the limited information of channel path directions and gains, that is, $\{\theta_l, \phi_l, \theta_{l,{\rm r}}, \phi_{l,{\rm r}}\}_{l=1}^{L_{\rm B}}$, $\{\theta_{i,{\rm t}}, \phi_{i,{\rm t}}\}_{i=1}^{L_{\rm M}}$, and ${\bf g}_{\rm B}\otimes {\bf g}_{\rm M}$. On the one hand, the overhead is dominated by the feedback of path gains because the path directions change slowly so that the need for their feedback is not as frequent as that for path gains. On the other hand, although channel reciprocity does not exist in FDD systems, spatial reciprocity holds for the BS--RIS and RIS--MS channels. Therefore, following the channel reconstruction in \cite{Han_FDD}, $\{\theta_l, \phi_l, \theta_{l,{\rm r}}, \phi_{l,{\rm r}}\}_{l=1}^{L_{\rm B}}$ in $\bf H$ and $\{\theta_{i,{\rm t}}, \phi_{i,{\rm t}}\}_{i=1}^{L_{\rm M}}$ in $\bf h$ can be extracted from the uplink and utilized to construct the downlink channel even without sending the path directions back to the BS. Given the above observations, we focus on the feedback of path gains ${\bf g}_{\rm B}\otimes {\bf g}_{\rm M}$ and assume that the perfect path directions are available at the BS.\par

Assume that the MS utilizes total $b$ bits for feedback. Since path gains in the BS--RIS and RIS--MS channels have different characteristic, $b_{\rm B}$ and $b_{\rm M}$ bits are allocated respectively to generate sub-codebooks for ${\bf g}_{\rm B}$ and ${\bf g}_{\rm M}$, where $b_{\rm B}+b_{\rm M}=b$. Furthermore, in the segmented channel, the NLoS path gains are fickle and independent identically distributed while the LoS path gain is relatively stable, which means they should not be quantized in the same way. Therefore, we will quantize the LoS and NLoS path gains of the BS--RIS channel with $b_{\rm B,L}$ and $b_{\rm B,N}$ bits\footnote{Although the NLoS path gains change faster than the LoS path gain, they are coupled in the MRT design. To reduce the rate loss caused by quantization error, they should be simultaneously quantized, as detailed in the following section.}, respectively, where $b_{\rm B,L}+b_{\rm B,N}=b_{\rm B}$. Similarly, $b_{\rm M}$ will be partitioned into $b_{\rm M,L}$ and $b_{\rm M,N}$ for the RIS--MS channel. On the basis of bit partitioning $b=b_{\rm B,L}+b_{\rm B,N}+b_{\rm M,L}+b_{\rm M,N}$, we design a cascaded codebook to quantize the path gain direction vector ${\tilde{\bf{ g}}_{\rm{B}}} \otimes {\tilde{\bf{ g}}_{\rm{M}}} = {{\bf{g}}_{\rm{B}}} \otimes {{\bf{g}}_{\rm{M}}}/\left\| {{{\bf{g}}_{\rm{B}}} \otimes {{\bf{g}}_{\rm{M}}}} \right\|$.\par

The cascaded codebook for quantization is given by
\begin{equation}\label{eq:W}
  {\mathcal W} = \left\{ {{{\bf{w}}_1},{{\bf{w}}_2}, \cdots ,{{\bf{w}}_{{{\rm{2}}^b}}}} \right\},
\end{equation}
where the codeword ${{\bf{w}}_i} \in {{\mathbb C}^{{L_{\rm{B}}}{L_{\rm{M}}} \times 1}}$ satisfies $\left\| {{{\bf{w}}_i}} \right\| = 1$. With the codebook $\mathcal W$, the path gain direction vector ${\tilde{\bf{ g}}_{\rm{B}}} \otimes {\tilde{\bf{ g}}_{\rm{M}}}$ can be quantized by ${\bf w}_{i^\#}$, where index $i^\# $ is obtained by
\begin{equation}\label{eq:i}
  {i^\# } = \mathop {\arg\max }\limits_{i \in \left\{ {1,2, \cdots ,{2^b}} \right\}} {\left| {{{\left( {{{\tilde{\bf{ g}}}_{\rm{B}}} \otimes {{\tilde{\bf{ g}}}_{\rm{M}}}} \right)}^H}{{\bf{w}}_i}} \right|^2}.
\end{equation}
As ${\tilde{\bf{ g}}_{\rm{B}}}$ and ${\tilde{\bf{ g}}_{\rm{M}}}$ are coupled with the Kronecker product, we also design the codeword ${\bf w}_i$ as the Kronecker product of two sub-codewords. With feedback bits $b_{\rm B}$ and $b_{\rm M}$, the sub-codebooks for the BS--RIS and RIS--MS channels are given respectively as ${{\mathcal W}_{\rm{B}}} = \{ {{{\bf{w}}_{{\rm{B}},1}},{{\bf{w}}_{{\rm{B}},2}}, \cdots ,{{\bf{w}}_{{\rm{B}},{2^{{b_{\rm{B}}}}}}}} \}$ and ${{\mathcal W}_{\rm{M}}} = \{ {{{\bf{w}}_{{\rm{M}},1}},{{\bf{w}}_{{\rm{M}},2}}, \cdots ,{{\bf{w}}_{{\rm{M}},{2^{{b_{\rm{M}}}}}}}} \}
$,
where ${{\bf{w}}_{{\rm{B}},{i_{\rm{B}}}}} \in {{\mathbb C}^{{L_{\rm{B}}} \times 1}}$ satisfying $\left\| {{{\bf{w}}_{{\rm{B}},{i_{\rm{B}}}}}} \right\| = 1$ and ${{\bf{w}}_{{\rm{M}},{i_{\rm{M}}}}} \in {{\mathbb C}^{{L_{\rm{M}}} \times 1}}$ satisfying $\left\| {{{\bf{w}}_{{\rm{M}},{i_{\rm{M}}}}}} \right\| = 1$ are sub-codewords. At this point, the codeword ${\bf w}_{i}$ can be synthesized as ${{\bf{w}}_i} = {{\bf{w}}_{{\rm{B}},{i_{\rm{B}}}}} \otimes {{\bf{w}}_{{\rm{M}},{i_{\rm{M}}}}}$, where ${i_{\rm{B}}} = {v_{{2^{{b_{\rm{M}}}}}}}\left( i \right)$ and ${i_{\rm{M}}} = {h_{{2^{{b_{\rm{M}}}}}}}\left( i \right)$. Next, the maximizing problem in \eqref{eq:i} can be rewritten as
\begin{equation}\label{eq:i_1}
  \begin{aligned}
 &\mathop {\max }\limits_{i \in \left\{ {1,2, \cdots ,{2^b}} \right\}} {\left| {\tilde{\bf{ g}}_{\rm{B}}^H{{\bf{w}}_{{\rm{B}},{i_{\rm{B}}}}}} \right|^2}{\left| {\tilde{\bf{ g}}_{\rm{M}}^H{{\bf{w}}_{{\rm{M}},{i_{\rm{M}}}}}} \right|^2}=  \left(\mathop {\max }\limits_{{i_{\rm B}} \in \left\{ {1,2, \cdots ,{2^{{b_{\rm{B}}}}}} \right\}} {\left| {\tilde{\bf{ g}}_{\rm{B}}^H{{\bf{w}}_{{\rm{B}},{i_{\rm{B}}}}}} \right|^2}\right) \times \left( \mathop {\max }\limits_{{i_{\rm M}} \in \left\{ {1,2, \cdots ,{2^{{b_{\rm{M}}}}}} \right\}} {\left| {\tilde{\bf{ g}}_{\rm{M}}^H{{\bf{w}}_{{\rm{M}},{i_{\rm{M}}}}}} \right|^2}\right).
\end{aligned}
\end{equation}
%\begin{equation}\label{eq:i_1}
%  \begin{aligned}
%{i^\# } &= \arg \mathop {\max }\limits_{i \in \left\{ {1,2, \cdots ,{2^b}} \right\}} {\left| {{\bf{\tilde g}}_{\rm{B}}^H{{\bf{w}}_{{\rm{B}},{i_{\rm{B}}}}}} \right|^2}{\left| {{\bf{\tilde g}}_{\rm{M}}^H{{\bf{w}}_{{\rm{M}},{i_{\rm{M}}}}}} \right|^2}\\
% &= \arg \mathop {\max }\limits_{{i_{\rm B}} \in \left\{ {1,2, \cdots ,{2^{{b_{\rm{B}}}}}} \right\}} {\left| {{\bf{\tilde g}}_{\rm{B}}^H{{\bf{w}}_{{\rm{B}},{i_{\rm{B}}}}}} \right|^2} \times \arg \mathop {\max }\limits_{{i_{\rm M}} \in \left\{ {1,2, \cdots ,{2^{{b_{\rm{M}}}}}} \right\}} {\left| {{\bf{\tilde g}}_{\rm{M}}^H{{\bf{w}}_{{\rm{M}},{i_{\rm{M}}}}}} \right|^2}
%\end{aligned}
%\end{equation}
The quantization procedure for $\tilde{\bf{ g}}_{\rm{B}}$ and $\tilde{\bf{ g}}_{\rm{M}}$ can be analyzed independently. Thus, ${{\bf{w}}_{{\rm{B}},{i_{\rm{B}}}}}$ and ${{\bf{w}}_{{\rm{M}},{i_{\rm{M}}}}}$ will be separately designed in the same way. We take the BS--RIS channel path gains as an example to illustrate our cascaded codebook design and feedback scheme.\par
With the sub-codebook ${{\mathcal W}_{\rm{B}}}$, ${\tilde{\bf{ g}}_{\rm{B}}}$ can be quantized as ${{\bf{w}}_{{\rm{B}},i_{\rm{B}}^\# }}$, where index $i_{\rm{B}}^\# $ is obtained by
\begin{equation}\label{eq:i_B}
  i_{\rm{B}}^\#  = \mathop {\arg \max }\limits_{{i_{\rm{B}}} \in \left\{ {1,2, \cdots ,{2^{{b_{\rm{B}}}}}} \right\}} {\left| {\tilde{\bf{ g}}_{\rm{B}}^H{{\bf{w}}_{{\rm{B}},{i_{\rm{B}}}}}} \right|^2}.
\end{equation}
When the energy and distribution differences of the LoS and NLoS path gains are considered, the cascaded codeword ${{\bf{w}}_{{\rm{B}},{i_{\rm{B}}}}}$ is designed as
\begin{equation}\label{eq:cas_codeword}
  {{\bf{w}}_{{\rm{B}},{i_{\rm{B}}}}} = {\left[ {\sqrt {\frac{{{K_{\rm{B}}}}}{{{K_{\rm{B}}} + 1}}} {w_{{\rm{B,L}},{i_{{\rm{B,L}}}}}},\sqrt {\frac{1}{{{K_{\rm{B}}} + 1}}} {\bf{w}}_{{\rm{B,N}},{i_{{\rm{B,N}}}}}^T} \right]^T},
\end{equation}
where ${w_{{\rm{B,L}},{i_{{\rm{B,L}}}}}}$ and ${{\bf{w}}_{{\rm{B,N}},{i_{{\rm{B,N}}}}}}$ are used to quantize ${{\tilde g}_{{\rm{B,1}}}}$ and ${\tilde{\bf{ g}}_{{\rm{B,N}}}} = {\left[ {{{\tilde g}_{{\rm{B,2}}}},{{\tilde g}_{{\rm{B,2}}}}, \cdots ,{{\tilde g}_{{\rm{B,}}{L_{\rm{B}}}}}} \right]^T}$, respectively. We first consider the quantization of ${\tilde{\bf{ g}}_{{\rm{B,N}}}}$. Given that the NLoS path gains vector is Rayleigh distributed, ${{\bf{w}}_{{\rm{B,N}},{i_{{\rm{B,N}}}}}}$ is selected from a size $2^{b_{\rm B,N}}$ RVQ codebook ${{\mathcal W}_{{\rm{B,N}}}}$, whose codeword is independent and isotropically distributed on a complex unit $L_{\rm B}-1$ dimensional hypersphere. With ${{\mathcal W}_{{\rm{B,N}}}}$, ${\tilde{\bf{ g}}_{{\rm{B,N}}}}$ can be quantized as ${{\bf{w}}_{{\rm{B,N}},i_{{\rm{B,N}}}^\# }}$, where the index $i_{{\rm{B,N}}}^\# $ is given by
\begin{equation}\label{eq:i_B,N}
  i_{{\rm{B,N}}}^\#  = \mathop {\arg\max }\limits_{{i_{{\rm{B,N}}}} \in \left\{ {1,2, \cdots ,{2^{{b_{{\rm{B,N}}}}}}} \right\}} {\left| {\tilde{\bf{ g}}_{{\rm{B,N}}}^H{{\bf{w}}_{{\rm{B,N}},{i_{{\rm{B,N}}}}}}} \right|^2}.
\end{equation}
Next, to achieve the maximal ${| {\tilde{\bf{ g}}_{\rm{B}}^H{{\bf{w}}_{{\rm{B}},{i_{\rm{B}}}}}} |^2}$, ${w_{{\rm{B,L}},{i_{{\rm{B,L}}}}}} = {e^{j{\varepsilon _{{i_{{\rm{B,L}}}}}}}}$ is designed, where ${\varepsilon _{{i_{{\rm{B,L}}}}}} ={i_{{\rm{B,L}}}} \frac{{2\pi }}{{{2^{{b_{{\rm{B,L}}}}}}}}$ is the quantized phase to align the phase of ${\tilde g_{{\rm{B,1}}}^H{w_{{\rm{B,L}},{i_{{\rm{B,L}}}}}}}$ and ${\tilde{\bf{ g}}_{{\rm{B,N}}}^H{{\bf{w}}_{{\rm{B,N}},i_{{\rm{B,N}}}^\# }}}$. Assuming ${\tilde g_{{\rm{B,1}}}}$ is optimally quantized as ${w_{{\rm{B,L}},i_{{\rm{B,L}}}^\# }}$, then the index $i_{{\rm{B,L}}}^\#$ should be derived by
\begin{equation}\label{eq:i_B,L}
  i_{{\rm{B,L}}}^\#  = \mathop {\arg\max }\limits_{{i_{{\rm{B,L}}}} \in \left\{ {1,2, \cdots ,{2^{{b_{{\rm{B,L}}}}}}} \right\}} {\left| {\sqrt {\frac{{{K_{\rm{B}}}}}{{{K_{\rm{B}}} + 1}}} \tilde g_{{\rm{B,1}}}^H{w_{{\rm{B,L}},{i_{{\rm{B,L}}}}}} + \sqrt {\frac{1}{{{K_{\rm{B}}} + 1}}} \tilde{\bf{ g}}_{{\rm{B,N}}}^H{{\bf{w}}_{{\rm{B,N}},i_{{\rm{B,N}}}^\# }}} \right|^2}.
\end{equation}
Obviously, the quantization procedures \eqref{eq:i_B,N} and \eqref{eq:i_B,L} are equivalent with \eqref{eq:i_B}. Therefore, we have $i_{\rm{B}}^\#  = i_{{\rm{B,L}}}^\#  + {2^{{b_{{\rm{B,L}}}}}}( {i_{{\rm{B,N}}}^\#  - 1} )$. As $b_{\rm B,L}$ is limited, the quantization error in \eqref{eq:i_B,L} satisfies
\begin{equation}\label{eq:phase_error}
  \left| {\angle \tilde g_{{\rm{B,1}}}^H{w_{{\rm{B,L}},i_{{\rm{B,L}}}^\# }} - \angle \tilde{\bf{ g}}_{{\rm{B,N}}}^H{{\bf{w}}_{{\rm{B,N}},i_{{\rm{B,N}}}^\# }}} \right| < {\pi }/{{{2^{{b_{{\rm{B,L}}}}}}}}.
\end{equation}\par
The above cascaded sub-codebook design and quantization procedure for $\tilde {\bf g}_{{\rm{B}}}$ are also applicable to $\tilde {\bf g}_{{\rm{M}}}$. After $i_{{\rm{B}}}^\#$ and $i_{{\rm{M}}}^\#$ are obtained, the optimal codeword index $i^\# = i_{{\rm{M}}}^\# + 2^{b_{\rm M}} (i_{{\rm{B}}}^\#-1)$ is fed back to the BS to reconstruct the quantized downlink channel as
\begin{equation}\label{eq:quan_h_eff}
  {{{\bf{\hat h}}}_{{\rm{eff}}}} = {Q_1}\left\| {{{\bf{g}}_{\rm{B}}} \otimes {{\bf{g}}_{\rm{M}}}} \right\|{\bf{A}} {{{\bf{w}}_{i^\# }} }.
\end{equation}
Note that we leave out the feedback for $\left\| {{{\bf{g}}_{\rm{B}}} \otimes {{\bf{g}}_{\rm{M}}}} \right\|$ because it is a constant scalar that requires negligible feedback overhead and can be absorbed into $Q_1$ . When the MRT is adopted at the BS, the constant scalar vanishes as described in the next section.

\subsection{Ergodic Rate Loss Analysis}\label{sec:4.1}
According to \eqref{eq:Jensen}, the upper bound on ergodic rate under quantized CSI is given by
\begin{equation}\label{eq:Jensn_2}
  {R_{{\rm{Q,upper}}}}= {\log _2}\left( {1 + {\hat P}/{{{\sigma ^2}}}} \right) ,
\end{equation}
where $\hat P={{\mathbb E}\{ {{{| {{\bf{h}}_{{\rm{eff}}}^H{\bf{\hat f}}} |}^2}} \}}$ is the expected received signal power with MRT ${\bf{\hat f}}$ designed utilizing quantized CSI. The difference of the upper bound ergodic rate is defined as ergodic rate loss, that is,
\begin{equation}\label{eq:rate_gap}
  \Delta R = {R_{{\rm{P,upper}}}} - {R_{{\rm{Q,upper}}}} = {\log _2}\left( {\frac{{{\sigma ^2} + {P_{{\rm{opt}}}}}}{{{\sigma ^2} + \hat P}}} \right).
\end{equation}
To have direct cognition on the ergodic rate loss, the analytical expression of $\hat P$ is imperative. In $\hat P$, the MRT designed with quantized CSI is expressed as
\begin{equation}\label{eq:quan_f}
  {\bf{\hat f}} = {{{{{\bf{\hat h}}}_{{\rm{eff}}}}}}/{{\left\| {{{{\bf{\hat h}}}_{{\rm{eff}}}}} \right\|}} = {{{\bf{A}}{{\bf{w}}_{{i^\# }}}}}/{{\left\| {{\bf{A}}{{\bf{w}}_{{i^\# }}}} \right\|}}.
\end{equation}
Note that ${{\left\| {{\bf{A}}{{\bf{w}}_{{i^\# }}}} \right\|}}$ is intractable during the analysis procedure. Thus its property is studied before we move on. With regard to the cascaded path direction matrix ${\bf A}$, we have the following property.
\begin{property}\label{Them:1}
For systems equipped with large-scale UPA at both the BS and the RIS (i.e., ${N_{\rm{B}}}\rightarrow\infty$ and ${N_{\rm{R}}}\rightarrow\infty$), when reflection coefficients of the RIS are designed to align the LoS links of the BS--RIS and RIS--MS channels, the cascaded path direction matrix of the RIS-assisted system satisfies
\begin{equation}\label{eq:them_1}
  {{\bf{A}}^H}{\bf{A}} = \frac{1}{{{N_{\rm{R}}}}}{{\bf{I}}_{{L_{\rm M}L_{\rm B}} \times {L_{\rm M}L_{\rm B}}}} + {{\bf{O}}_1},
\end{equation}
where
\begin{equation}\label{eq:O_opt}
  {{\bf{O}}_1} = \left[ {\begin{array}{*{20}{c}}
{1 - \frac{1}{{{N_{\rm{R}}}}}}&{{{\bf{0}}_{1 \times \left( {{L_{\rm M}L_{\rm B}} - 1} \right)}}}\\
{{{\bf{0}}_{\left( {{L_{\rm M}L_{\rm B}} - 1} \right) \times 1}}}&{{{\bf{0}}_{\left( {{L_{\rm M}L_{\rm B}} - 1} \right) \times \left( {{L_{\rm M}L_{\rm B}} - 1} \right)}}}
\end{array}} \right].
\end{equation}
\end{property}
\begin{IEEEproof}
See Appendix \ref{App:A}.
\end{IEEEproof}
\emph{Property \ref{Them:1}} shows that $L_{\rm B}L_{\rm M}$ paths cascaded by RIS are asymptotically orthogonal to one another, and designing the reflection coefficients for a specific cascaded path can dramatically increase its energy. On the basis of \emph{Property \ref{Them:1}}, we have\footnote{For clear and concise description, we omit the codeword indices.}
\begin{equation}\label{eq:Q_2}
  \begin{aligned}
\frac{1}{{Q_2^2}}&\buildrel \Delta \over ={{\left\| {{\bf{A}}{{\bf{w}}}} \right\|^2}}= {\left( {{{\bf{w}}_{\rm{B}}} \otimes {{\bf{w}}_{\rm{M}}}} \right)^H}\left( {\frac{1}{{{N_{\rm{R}}}}}{\bf{I + }}{{\bf{O}}_1}} \right)\left( {{{\bf{w}}_{\rm{B}}} \otimes {{\bf{w}}_{\rm{M}}}} \right)\\
 &= \frac{1}{{{N_{\rm{R}}}}}{\bf{w}}_{\rm{B}}^H{{\bf{w}}_{\rm{B}}}{\bf{w}}_{\rm{M}}^H{{\bf{w}}_{\rm{M}}}{\bf{ + }}{\left| {\sqrt {\frac{{{K_{\rm{B}}}}}{{{K_{\rm{B}}} + 1}}} {e^{j{\varepsilon _{\rm{B}}}}}} \right|^2}{\left| {\sqrt {\frac{{{K_{\rm{M}}}}}{{{K_{\rm{M}}} + 1}}} {e^{j{\varepsilon _{\rm{M}}}}}} \right|^2}\left( {1 - \frac{1}{{{N_{\rm{R}}}}}} \right)\\
 &= \frac{1}{{{N_{\rm{R}}}}}{\bf{ + }}\frac{{{K_{\rm{B}}}{K_{\rm{M}}}\left( {{N_{\rm{R}}} - 1} \right)}}{{\left( {{K_{\rm{B}}} + 1} \right)\left( {{K_{\rm{M}}} + 1} \right){N_{\rm{R}}}}}.
\end{aligned}
\end{equation}
Now, \eqref{eq:quan_f} can be simplified as
\begin{equation}\label{eq:f_simp}
  {\bf{\hat f}} = {Q_2}{\bf{A}}\left( {{{\bf{w}}_{{\rm{B}},i_{\rm{B}}^\# }} \otimes {{\bf{w}}_{{\rm{M}},i_{\rm{M}}^\# }}} \right).
\end{equation}
%When the quantized MRT in \eqref{eq:f_simp} and optimal reflection phase in \eqref{eq:optimal_phase} are designed, we apply the Jensen's inequality again to the ergodic rate
%\begin{equation}\label{eq:Jensn_2}
%  {R_Q} = {\mathbb E}\left\{ {{{\log }_2}\left( {1 + \frac{{{{\left| {{\bf{h}}_{{\rm{eff}}}^H{\bf{\hat f}}} \right|}^2}}}{{{\sigma ^2}}}} \right)} \right\} \le {\log _2}\left( {1 + \frac{\hat P}{{{\sigma ^2}}}} \right) = {R_{{\rm{Q,upper}}}},
%\end{equation}
%where $\hat P={{\mathbb E}\left\{ {{{\left| {{\bf{h}}_{{\rm{eff}}}^H{\bf{\hat f}}} \right|}^2}} \right\}}$ is the expected received signal power under quantized CSI. The difference of upper bound ergodic rate is defined as ergodic rate loss, i.e.,
%\begin{equation}\label{eq:rate_gap}
%  \Delta R = {R_{{\rm{P,upper}}}} - {R_{{\rm{Q,upper}}}} = {\log _2}\left( {\frac{{{\sigma ^2} + {P_{{\rm{opt}}}}}}{{{\sigma ^2} + \hat P}}} \right)
%\end{equation}
%To have a direct cognition on the ergodic rate loss, the analytical expression of $\hat P$ is an imperative.\par
Substituting \eqref{eq:h_eff} and \eqref{eq:f_simp} into $\hat P={{\mathbb E}\{ {{{| {{\bf{h}}_{{\rm{eff}}}^H{\bf{\hat f}}} |}^2}} \}}$, we have
\begin{equation}\label{eq:P_expand}
  \begin{aligned}
\hat P = Q_1^2Q_2^2{\mathbb E}\left\{ {{{\left| {\underbrace {\frac{1}{{{N_{\rm{R}}}}}{{\left( {{{\bf{g}}_{\rm{B}}} \otimes {{\bf{g}}_{\rm{M}}}} \right)}^H}\left( {{{\bf{w}}_{{\rm{B}},i_{\rm{B}}^\# }} \otimes {{\bf{w}}_{{\rm{M}},i_{\rm{M}}^\# }}} \right)}_{{X_1}} + \underbrace {{{\left( {{{\bf{g}}_{\rm{B}}} \otimes {{\bf{g}}_{\rm{M}}}} \right)}^H}{{\bf{O}}_{1}}\left( {{{\bf{w}}_{{\rm{B}},i_{\rm{B}}^\# }} \otimes {{\bf{w}}_{{\rm{M}},i_{\rm{M}}^\# }}} \right)}_{{X_2}}} \right|}^2}} \right\}.
\end{aligned}
\end{equation}
According to the cosine theorem, the expectation in \eqref{eq:P_expand} can be expanded as
\begin{equation}\label{eq:triangle}
  {\mathbb E}\left\{ {{{\left| {{X_1} + {X_2}} \right|}^2}} \right\} = {\mathbb E}\left\{ {{{\left| {{X_1}} \right|}^2}} \right\} + {\mathbb E}\left\{ {{{\left| {{X_2}} \right|}^2}} \right\} - 2{\mathbb E}\left\{ {\left| {{X_1}} \right|\left| {{X_2}} \right|\cos \varpi } \right\},
\end{equation}
where $\varpi$ is the supplementary angle of the angle between $X_1$ and $X_2$, which is given by
\begin{equation}\label{eq:angle_cos}
  \varpi  = \pi  - \left| {\angle {X_1} - \angle {X_2}} \right|.
\end{equation}
In \eqref{eq:triangle}, ${\mathbb E}\{ {{{| {{X_2}} |}^2}} \}$ is the expected received signal power contributed by the cascaded LoS path and has an analytical expression that is shown in \emph{Lemma \ref{lemma:1}}.
\begin{lemma}\label{lemma:1}
In the RIS-assisted system, when the MRT and optimal reflection phase design \eqref{eq:optimal_phase} are adopted at the BS and RIS, respectively, the expected received signal power contributed by the cascaded LoS channel can be expressed as
\begin{equation}\label{eq:E_X_2}
  {Q_3} \buildrel \Delta \over = {\mathbb E}\left\{ {{{\left| {{X_2}} \right|}^2}} \right\} = \frac{{K_{\rm{B}}^2K_{\rm{M}}^2{{\left( {{N_{\rm{R}}} - 1} \right)}^2}\left( {{L_{\rm{B}}} - 1} \right)\left( {{L_{\rm{M}}} - 1} \right)}}{N_{\rm{R}}^2{\left( {{K_{\rm{B}}} + 1} \right)\left( {{K_{\rm{M}}} + 1} \right)}}.
\end{equation}
\end{lemma}
\begin{IEEEproof}
See Appendix \ref{App:lemma-1}.
\end{IEEEproof}
It can be seen from \eqref{eq:angle_cos} that $\varpi$ is determined by the phase of $X_1$ and $X_2$, not their amplitude; thus, ${\mathbb E}\{ {| {{X_1}} || {{X_2}} |\cos \varpi } \} = {\mathbb E}\{ {| {{X_1}} || {{X_2}} | } \}{\mathbb E}\{ {\cos \varpi } \}$. To further explore $\hat P$, we approximate ${\mathbb E}\{ {| {{X_1}} || {{X_2}} | } \}$ in \emph{Lemma \ref{Lemma:2}}.
\begin{lemma}\label{Lemma:2}
When Rician factors are sufficiently large, ${\mathbb E}\{ {| {{X_1}} || {{X_2}} |} \}$ can be approximated by
\begin{equation}\label{eq:E_X_1_X_2}
  {Q_4} \buildrel \Delta \over = {\mathbb E}\left\{ {\left| {{X_1}} \right|\left| {{X_2}} \right|} \right\} \approx {Q_3}/( {{N_{\rm{R}}} - 1} ).
\end{equation}
\end{lemma}
\begin{IEEEproof}
See Appendix \ref{App:E}.
\end{IEEEproof}
With the application of \emph{Lemma \ref{lemma:1}} and \emph{Lemma \ref{Lemma:2}}, the expected received signal power under quantized CSI can be expressed as
\begin{equation}\label{eq:P_bef_Q}
  \hat P \approx Q_1^2Q_2^2\left( {{\mathbb E}\left\{ {{{\left| {{X_1}} \right|}^2}} \right\} + {Q_3} + 2{Q_4}{\mathbb E}\left\{-\cos \varpi \right\}} \right),
\end{equation}
where ${\mathbb E}\{ {{{| {{X_1}} |}^2}} \}$ and ${\mathbb E}\{-\cos \varpi \}$ are associated with the quantization error determined by the total feedback bits. In next subsection, the quantization error will be analyzed to derive a closed-form lower bound on $\hat P$.
\subsection{Quantization Error Analysis}\label{sec:4.2}
In practice, feedback bits for quantizing path gains are limited, which will inevitably incur quantization error. Consider that the total feedback bits are divided to design different parts of the proposed cascaded codebook. The effects of different feedback bits on the quantization error are exhibited in the following two theorems to guide the bit partitioning strategy.

\begin{theorem}\label{Them:3}
On the basis of the proposed cascaded codebook, when LoS path gains in the BS--RIS and RIS--MS channels are quantized with $b_{\rm B,L}$ and $b_{\rm M,L}$ bits, the angle between $X_1$ and $X_2$, $\varpi$, satisfies
\begin{equation}\label{eq:Them-3-1}
  \pi  - \left( {\frac{\pi }{{{2^{{b_{{\rm{B,L}}}}}}}} + \frac{\pi }{{{2^{{b_{{\rm{M,L}}}}}}}}} \right) < \varpi  \leq \pi.
\end{equation}
This produces a lower bound for ${\mathbb E}\left\{-\cos \varpi \right\}$ as
\begin{equation}\label{eq:Them-3-2}
  {\mathbb E}\left\{-\cos \varpi \right\} > {\cos \left( {\frac{\pi }{{{2^{{b_{{\rm{B,L}}}}}}}} + \frac{\pi }{{{2^{{b_{{\rm{M,L}}}}}}}}} \right)}.
\end{equation}
\end{theorem}
\begin{IEEEproof}
See Appendix \ref{App:G}.
\end{IEEEproof}

\begin{theorem}\label{Them:2}
On the basis of the proposed cascaded codebook, the averaged inner product between the path gain vector and the optimal codeword has a lower bound, that is,
\begin{equation}\label{eq:Them-2-1}
  {\mathbb E}\left\{ {{{\left| {{X_1}} \right|}^2}} \right\} > {Q_5}\left( {\kappa_{\rm B} - {2^{\frac{{ - {b_{{\rm{B,N}}}}}}{{{L_{\rm{B}}} - 2}}}} } \right)\left( {\kappa_{\rm M} - {2^{\frac{{ - {b_{{\rm{M,N}}}}}}{{{L_{\rm{M}}} - 2}}}} } \right),
\end{equation}
where ${Q_5} = {{\left( {{L_{\rm{B}}} - 1} \right)\left( {{L_{\rm{M}}} - 1} \right)}}/{{N_{\rm{R}}^2}}$, $\kappa_{\rm B}={{K_{\rm{B}}^2}}/({{{K_{\rm{B}}} + 1}}) +  1$, and $\kappa_{\rm M}={{K_{\rm{M}}^2}}/({{{K_{\rm{M}}} + 1}} )+  1$.
\end{theorem}
\begin{IEEEproof}
See Appendix \ref{App:F}.
\end{IEEEproof}
\emph{Theorems \ref{Them:3}} and \emph{\ref{Them:2}} uncover that feedback bits for LoS and NLoS affect the expected received signal power in a different way. As mentioned in \eqref{eq:i_B,L}, the role of feedback bits for LoS path gains is to align phases of the LoS and NLoS components; therefore, $b_{\rm B,L}$ and $b_{\rm M,L}$ jointly determine a lower bound on the angle between $X_1$ and $X_2$. Moreover, as we discuss in Appendix \ref{App:F}, ${\mathbb E}\{ {{{| {{X_1}} |}^2}} \}$ depends exclusively on the quantization error brought by using the RVQ codebook and is thus irrelevant to $b_{\rm B,L}$ and $b_{\rm M,L}$. This separated property benefits the bit partitioning strategy in Section \ref{sec:5}. By substituting \eqref{eq:Them-2-1} and \eqref{eq:Them-3-2} into \eqref{eq:P_bef_Q}, we derive a lower bound on the expected received signal power under quantized CSI as
\begin{equation}\label{eq:P_lower}
\begin{aligned}
  \hat P_{\rm lower} = Q_1^2Q_2^2\left( {{Q_5}\left( {\kappa_{\rm B} - {2^{\frac{{ - {b_{{\rm{B,N}}}}}}{{{L_{\rm{B}}} - 2}}}} } \right)\left( {\kappa_{\rm M} - {2^{\frac{{ - {b_{{\rm{M,N}}}}}}{{{L_{\rm{M}}} - 2}}}} } \right) + {Q_3} + 2{Q_4}\cos\left( {\frac{\pi }{{{2^{{b_{{\rm{B,L}}}}}}}} + \frac{\pi }{{{2^{{b_{{\rm{M,L}}}}}}}}} \right)} \right).
\end{aligned}
\end{equation}
Finally, the ergodic rate loss in \eqref{eq:rate_gap} can be upper bounded by
\begin{equation}\label{eq:det_R_upper}
  \Delta R < {\log _2}\left( {\frac{{{\sigma ^2} + {P_{{\rm{opt}}}}}}{{{\sigma ^2} + {{\hat P}_{{\rm{lower}}}}}}} \right) \buildrel \Delta \over = \Delta {R_{{\rm{upper}}}}.
\end{equation}
The closed-form expression of $\Delta {R_{{\rm{upper}}}}$ is a function of feedback bits $b_{\rm B,L}$, $b_{\rm M,L}$, $b_{\rm B,N}$, and $b_{\rm M,N}$. As the number of total feedback bits is fixed, it is possible to improve the ergodic rate by dividing feedback bits to separately quantize the LoS and NLoS path gains of both the BS--RIS and RIS--MS channels.
%%%%%%%%%%%%%%%%%%%%%%%%%%%%%%%%%%%%%%%%%%%%%%%%%%%%%%%%%%%%%%%%%%%%%%%%%%%%%%%%%%%%%%%%%%%%%%
\section{Adaptive Bit Partitioning}\label{sec:5}
In this section, we propose an adaptive bit partitioning strategy to minimize the upper bound on the ergodic rate loss. Specifically, assuming ${b_{\rm{L}}} \buildrel \Delta \over = {b_{{\rm{B,L}}}} + {b_{{\rm{M,L}}}}$ and ${b_{\rm{N}}} \buildrel \Delta \over = {b_{{\rm{B,N}}}} + {b_{{\rm{M,N}}}}$ are fixed, we investigate the bit allocation between LoS paths and between NLoS paths, respectively. Afterward, the segmentation for ${b_{\rm{L}}}$ and ${b_{\rm{N}}}$ is studied given $b = {b_{\rm{L}}} + {b_{\rm{N}}}$. Note that the adaptive bit partitioning strategy is inverse to the analysis process. Specifically, ${b_{\rm{L}}}$ and ${b_{\rm{N}}}$ are first partitioned given fixed $b$, and then the allocation goes to ${b_{{\rm{B,L}}}}$, ${b_{{\rm{M,L}}}}$, ${b_{{\rm{B,N}}}}$, and ${b_{{\rm{M,N}}}}$.
\subsection{Bit Partitioning for LoS Paths Between BS--RIS and RIS--MS}\label{sec:5.1}
The problem of minimizing the upper bound on ergodic rate loss is equivalent to maximizing $\hat P_{\rm lower}$. It is intractable to directly obtain closed-form expressions of the optimal ${b_{{\rm{B,L}}}}$, ${b_{{\rm{M,L}}}}$, ${b_{{\rm{B,N}}}}$, and ${b_{{\rm{M,N}}}}$ for maximizing $\hat P_{\rm lower}$ with fixed total feedback bits $b$. Fortunately, we have shown in \emph{Theorems \ref{Them:3}} and \emph{\ref{Them:2}} that feedback bits for LoS and NLoS quantization affect the expected received signal power in different ways. This offers us the idea of initially optimizing ${b_{{\rm{B,L}}}}$, ${b_{{\rm{M,L}}}}$, ${b_{{\rm{B,N}}}}$, and ${b_{{\rm{M,N}}}}$ with given ${b_{\rm{N}}}$ and ${b_{\rm{L}}}$ to maximize different parts of $\hat P_{\rm lower}$ in a separate way. We then split the limited $b$ into ${b_{\rm{N}}}$ and ${b_{\rm{L}}}$ to achieve the maximal $\hat P_{\rm lower}$. In this subsection, we focus on the bit partitioning between ${b_{{\rm{B,L}}}}$ and ${b_{{\rm{M,L}}}}$ when $b_{\rm L}$ is fixed.\par
The part of $\hat P_{\rm lower}$ that associates with ${b_{{\rm{B,L}}}}$ and ${b_{{\rm{M,L}}}}$ is denoted by the function
\begin{equation}\label{eq:P_lower_3}
  {\hat P_{{\rm{lower,3}}}}\left({b_{{\rm{B,L}}}}\right) \buildrel \Delta \over = 2{Q_4}\cos \left( {\frac{\pi }{{{2^{{b_{{\rm{B,L}}}}}}}} + \frac{\pi }{{{2^{{b_{{\rm{M,L}}}}}}}}} \right) = 2{Q_4}\cos \left( {\frac{{{2^{{b_{\rm{L}}} - {b_{{\rm{B,L}}}}}} + {2^{{b_{{\rm{B,L}}}}}}}}{{{2^{{b_{\rm{L}}}}}}}\pi } \right).
\end{equation}
For the sake of concise description, we define $g\left( {{b_{{\rm{B,L}}}}} \right) = {2^{{b_{\rm{L}}} - {b_{{\rm{B,L}}}}}} + {2^{{b_{{\rm{B,L}}}}}}$ and get its derivative as follows:
\begin{equation}\label{eq:de_f_b_B,L}
  g'\left( {{b_{{\rm{B,L}}}}} \right) = \ln 2\left( { - {2^{{b_{\rm{L}}} - {b_{{\rm{B,L}}}}}} + {2^{{b_{{\rm{B,L}}}}}}} \right).
\end{equation}
By letting $g'\left( {{b_{{\rm{B,L}}}}} \right) = 0$, we can obtain the optimal ${{b_{{\rm{B,L}}}}}$ that achieves minimal $g\left( {{b_{{\rm{B,L}}}}} \right)$ as
\begin{equation}\label{eq:b_B,L_opt}
  b_{{\rm{B,L}}}^{\dag} = {{{b_{\rm{L}}}}}/{2}.
\end{equation}
On the basis of \eqref{eq:b_B,L_opt}, the maximal ${\hat P_{{\rm{lower,3}}}}\left({b_{{\rm{B,L}}}}\right)$ is given by
\begin{equation}\label{eq:P_lower_3-max}
  \hat P_{{\rm{lower,3}}}^{\max } ={\hat P_{{\rm{lower,3}}}}\left(b_{{\rm{B,L}}}^{\dag}\right)= 2{Q_4}\cos \left( {\frac{\pi }{{\sqrt {{2^{{b_{\rm{L}}}{\rm{ - 2}}}}} }}} \right).
\end{equation}
Although LoS path gains in the BS--RIS and RIS--MS channels may vary in strength, the optimal ${b_{{\rm{B,L}}}}$ and ${b_{{\rm{M,L}}}}$, however, are equal to half of $b_{\rm L}$. We know that the ultimate aim of LoS path quantization is to align the phase of the LoS and NLoS components in \eqref{eq:i_B,L}, which is independent of amplitude. Hence, $b_{\rm L}$ should be equally split to ${b_{{\rm{B,L}}}}$ and ${b_{{\rm{M,L}}}}$ because the phases of two individual LoS path gains are combined equally in the cascaded LoS path. Clearly, the feedback bit allocation in \eqref{eq:b_B,L_opt} is in general not an integer. To meet the actual requirement, we modify the assignment as
\begin{equation}\label{eq:b_B,L_I}
  b_{{\rm{B,L}}}^{{\rm{\dag,I}}} = \left\{ \begin{aligned}
&\left\lceil {b_{{\rm{B,L}}}^{\rm{\dag}}} \right\rceil ,{\rm{ if }}\;{{\hat P}_{{\rm{lower,3}}}}\left( {\left\lceil {b_{{\rm{B,L}}}^{\rm{\dag}}} \right\rceil } \right) \ge {{\hat P}_{{\rm{lower,3}}}}\left( {\left\lfloor {b_{{\rm{B,L}}}^{\rm{\dag}}} \right\rfloor } \right)\\
&\left\lfloor {b_{{\rm{B,L}}}^{\rm{\dag}}} \right\rfloor ,{\rm{ otherwise}}
\end{aligned} \right..
\end{equation}
At this point, the bit partitioning strategy between LoS paths can be expressed as
\begin{equation}\label{eq:b4LoS}
  \left\{ \begin{aligned}
&b_{{\rm{B,L}}}^\#  = b_{{\rm{B,L}}}^{{\rm{\dag,I}}} \\
&b_{{\rm{M,L}}}^\#  = {b_{\rm{L}}} - b_{{\rm{B,L}}}^\#
\end{aligned} \right..
\end{equation}

\subsection{Bit Partitioning for NLoS Paths Between BS--RIS and RIS--MS}\label{sec:5.2}
Given $b_{\rm N}$ in this subsection, the allocation for ${b_{{\rm{B,N}}}}$ and ${b_{{\rm{M,N}}}}=b_{\rm N}-{b_{{\rm{B,N}}}}$ is analyzed to get an approximate maximum of the lower bound of ${\mathbb E}\{ {{{| {{X_1}} |}^2}} \}$, which is a part of $\hat P_{\rm lower}$. Insights on how the number of NLoS paths and Rician factors affect the bit partitioning criteria are also derived.\par
For the sake of convenience, we denote
\begin{equation}\label{eq:P_L_1}
  \begin{aligned}
{{\hat P}_{{\rm{lower,1}}}}\left({b_{{\rm{B,N}}}}\right) \buildrel \Delta \over = \left( {\kappa_{\rm B} - {2^{\frac{{ - {b_{{\rm{B,N}}}}}}{{{L_{\rm{B}}} - 2}}}}} \right)\left( {\kappa_{\rm M} - {2^{\frac{{ - {b_{{\rm{M,N}}}}}}{{{L_{\rm{M}}} - 2}}}}} \right) = {{\kappa_{\rm B}\kappa_{\rm M}}} - \kappa_{\rm B}{2^{\frac{{ - {b_{{\rm{M,N}}}}}}{{{L_{\rm{M}}} - 2}}}} - \kappa_{\rm M}{2^{\frac{{ - {b_{{\rm{B,N}}}}}}{{{L_{\rm{B}}} - 2}}}}+{2^{\frac{{ - {b_{{\rm{B,N}}}}}}{{{L_{\rm{B}}} - 2}}}}{2^{\frac{{ - {b_{{\rm{M,N}}}}}}{{{L_{\rm{M}}} - 2}}}}
\end{aligned}.
\end{equation}
Compared with the first third terms, ${2^{\frac{{ - {b_{{\rm{B,N}}}}}}{{{L_{\rm{B}}} - 2}}}}{2^{\frac{{ - {b_{{\rm{M,N}}}}}}{{{L_{\rm{M}}} - 2}}}}$ is negligible. Thus, by ignoring the last term in \eqref{eq:P_L_1}, we obtain the derivative of ${{\hat P}_{{\rm{lower,1}}}}\left({b_{{\rm{B,N}}}}\right)$ with respect to $b_{\rm B,N}$ as
\begin{equation}\label{eq:P_L_1_deri}
  {{\hat P}^\prime_{{\rm{lower,1}}}}\left({b_{{\rm{B,N}}}}\right) =  - \frac{{\kappa_{\rm B}\ln 2}}{{{L_{\rm{M}}} - 2}}{2^{ - \frac{{{b_{\rm{N}}} - {b_{{\rm{B,N}}}}}}{{{L_{\rm{M}}} - 2}}}} + \frac{{{\kappa_{\rm M}}\ln 2}}{{{L_{\rm{B}}} - 2}}{2^{ - \frac{{{b_{{\rm{B,N}}}}}}{{{L_{\rm{B}}} - 2}}}}.
\end{equation}
Letting ${{\hat P}^\prime_{{\rm{lower,1}}}}\left({b_{{\rm{B,N}}}}\right) =0$, the allocation for ${b_{{\rm{B,N}}}}$ can be given by
\begin{equation}\label{eq:b_B,N_opt}
  b_{{\rm{B,N}}}^{\dag} = \frac{{{b_{\rm{N}}}\left( {{L_{\rm{B}}} - 2} \right)}}{{\left( {{L_{\rm{B}}} + {L_{\rm{M}}} - 4} \right)}} + \frac{{\left( {{L_{\rm{M}}} - 2} \right)\left( {{L_{\rm{B}}} - 2} \right)}}{{\left( {{L_{\rm{B}}} + {L_{\rm{M}}} - 4} \right)}}{\log _2}\frac{{\kappa_{\rm M}\left( {{L_{\rm{M}}} - 2} \right)}}{{\kappa_{\rm B}\left( {{L_{\rm{B}}} - 2} \right)}},
\end{equation}
and the maximal ${\hat P_{{\rm{lower,1}}}}\left({b_{{\rm{B,N}}}}\right)$ is approximated by
\begin{equation}\label{eq:P_L_1_max}
  \hat P_{{\rm{lower,1}}}^{\max } = {\hat P_{{\rm{lower,1}}}}\left(b_{{\rm{B,N}}}^{\dag}\right) \approx {{\kappa_{\rm B}\kappa_{\rm M}}} - {Q_6}{2^{\frac{{ - {b_{\rm{N}}}}}{{{L_{\rm{B}}} + {L_{\rm{M}}} - 4}}}},
\end{equation}
where
\begin{equation}\label{eq:Q_6}
  {Q_6} = \kappa_{\rm B}{2^{\frac{{\left( {{L_{\rm{B}}} - 2} \right)}}{{\left( {{L_{\rm{B}}} + {L_{\rm{M}}} - 4} \right)}}{{\log }_2}\frac{{\kappa_{\rm M}\left( {{L_{\rm{M}}} - 2} \right)}}{{\kappa_{\rm B}\left( {{L_{\rm{B}}} - 2} \right)}}}} + \kappa_{\rm M}{2^{\frac{{ - \left( {{L_{\rm{M}}} - 2} \right)}}{{\left( {{L_{\rm{B}}} + {L_{\rm{M}}} - 4} \right)}}{{\log }_2}\frac{{\kappa_{\rm M}\left( {{L_{\rm{M}}} - 2} \right)}}{{\kappa_{\rm B}\left( {{L_{\rm{B}}} - 2} \right)}}}}.
\end{equation}
The closed-form of $b_{{\rm{B,N}}}^{\dag}$ reveals that the partitioning strategy between NLoS paths depends on the Rician factors and the number of NLoS paths. To figure out how these two factors determine the assignment process, we consider two special cases.\par
\subsubsection{Case 1}
The BS--RIS and RIS--MS channels have the same NLoS paths (i.e., ${L_{\rm{B}}} = {L_{\rm{M}}} = L$). In this case, \eqref{eq:b_B,N_opt} is simplified as
\begin{equation}\label{eq:b_B,N_opt_1}
  b_{{\rm{B,N}}}^{\dag}\Big {|}_ {{{\rm{case \,1}}}}  = \frac{{{b_{\rm{N}}}}}{2} + \frac{{L - 2}}{2}{\log _2}\frac{\kappa_{\rm M}}{\kappa_{\rm B}}.
\end{equation}
When path numbers are equal, we can see that the criterion of bit partitioning is to increase or decrease the feedback bits according to the Rician factors on the basis of equal segmentation $\frac{{{b_{\rm{N}}}}}{2}$. In addition, the step of increase or decrease is positively related to the number of paths. Specifically, if Rician factors are also the same, \eqref{eq:b_B,N_opt_1} degenerates into equal segmentation, and when ${K_{\rm{B}}} \ne {K_{\rm{M}}}$, by setting ${\log _2}\frac{\kappa_{\rm M}}{\kappa_{\rm B}} > 0$, we have ${K_{\rm{B}}} < {K_{\rm{M}}}$. This outcome implies that when the Rician factor of the BS--RIS channel is smaller than that of the RIS--MS channel, feedback bits assigned to the BS--RIS channel are larger than equal segmentation $\frac{{{b_{\rm{N}}}}}{2}$. Considering that Rician factor represents the energy ratio between the LoS and NLoS channel components, the channel with a smaller Rician factor owns a larger NLoS energy. Thus, it makes intuitive sense that the NLoS component of the BS--RIS channel contributes more to the system performance than that of the RIS--MS channel. Under this circumstance, more bits should be allocated to quantize NLoS paths in the BS--RIS channel for effective alleviation of the total quantization error.
\subsubsection{Case 2}
The BS--RIS and RIS--MS channels have the same Rician factor (i.e., ${K_{\rm{B}}} = {K_{\rm{M}}} = K$). In this scenario, feedback bits for ${\tilde{\bf{ g}}_{{\rm{B,N}}}}$ are given by
\begin{equation}\label{eq:b_B,N_opt_2}
  b_{{\rm{B,N}}}^{\dag}\Big{|}_ {{{\rm{case\, 2}}}}  = \frac{{{b_{\rm{N}}}\left( {{L_{\rm{B}}} - 2} \right)}}{{\left( {{L_{\rm{M}}} - 2} \right) + \left( {{L_{\rm{B}}} - 2} \right)}} + \frac{{\left( {{L_{\rm{M}}} - 2} \right)\left( {{L_{\rm{B}}} - 2} \right)}}{{\left( {{L_{\rm{M}}} - 2} \right) + \left( {{L_{\rm{B}}} - 2} \right)}}{\log _2}\frac{{\left( {{L_{\rm{M}}} - 2} \right)}}{{\left( {{L_{\rm{B}}} - 2} \right)}}.
\end{equation}
The first term in \eqref{eq:b_B,N_opt_2} shows that when $L_{\rm B}$ increases, more bits will be taken from $b_{\rm N}$ to quantize the NLoS paths in the BS--RIS channel. On the other hand, the second term plays a role in balancing the bit allocation criteria. For example, when ${L_{\rm{B}}} > {L_{\rm{M}}}$ and $L_{\rm B}$ keep increasing, the first term accounts for more than half of $b_{\rm N}$ while the second term will be negative to slow down the augmentation of $b_{\rm B,N}$ or even decrease it. The principle behind this is the tradeoff between quantization vector length and accuracy. When $b_{\rm N}$ is fixed and the BS--RIS channel has more path numbers, more bits should be allocated to $b_{\rm B,N}$ because the size of the RVQ codeword is larger. However, if the total energy of NLoS paths in the BS--RIS and RIS--MS channels are equal, then the individual NLoS path energy in the BS--RIS channel decreases. In this case, the required quantization accuracy for ${\tilde{\bf{ g}}_{{\rm{B,N}}}}$ is less important and $b_{\rm B,N}$ can be slightly reduced.\par
Owing to the integer limitation, we constrain the bit allocation in \eqref{eq:b_B,N_opt} as
\begin{equation}\label{eq:b_B,N_int}
  b_{{\rm{B,N}}}^{{\rm{\dag,I}}} = \left\{ \begin{aligned}
&\left\lceil {b_{{\rm{B,N}}}^{\rm{\dag}}} \right\rceil ,{\rm{ if }}\;{{\hat P}_{{\rm{lower,1}}}}\left( {\left\lceil {b_{{\rm{B,N}}}^{\rm{\dag}}} \right\rceil } \right) \ge {{\hat P}_{{\rm{lower,1}}}}\left( {\left\lfloor {b_{{\rm{B,N}}}^{\rm{\dag}}} \right\rfloor } \right)\\
&\left\lfloor {b_{{\rm{B,N}}}^{\rm{\dag}}} \right\rfloor ,{\rm{ otherwise}}
\end{aligned} \right..
\end{equation}
To avoid $b_{\rm B,N}$ going out of $[0,b_{\rm N}]$, the bit partitioning strategy between NLoS paths is expressed as
\begin{equation}\label{eq:b4NLoS}
  \left\{ \begin{aligned}
&b_{{\rm{B,N}}}^\#  = \min \left\{ {\max \left\{ {0, b_{{\rm{B,N}}}^{{\rm{\dag,I}}} } \right\},{b_{\rm{N}}}} \right\}\\
&b_{{\rm{M,N}}}^\#  = {b_{\rm{N}}} - b_{{\rm{B,N}}}^\#
\end{aligned} \right..
\end{equation}
\subsection{Bit Partitioning for LoS and NLoS Paths}\label{sec:5.3}
Thus far, we derived closed-form expressions for feedback bits ${b_{{\rm{B,L}}}}$, ${b_{{\rm{M,L}}}}$, ${b_{{\rm{B,N}}}}$, and ${b_{{\rm{M,N}}}}$ when $b_{\rm L}$ and $b_{\rm N}$ are fixed independently. Nevertheless, $b_{\rm L}$ and $b_{\rm N}$ are coupled through the limited total feedback bits $b$ and require proper allocation to achieve overall maximum of $\hat P_{\rm lower}$. On the basis of Sections \ref{sec:5.1} and \ref{sec:5.2}, when bit partitioning strategies in \eqref{eq:b_B,L_opt} and \eqref{eq:b_B,N_opt} are applied, ${\hat P_{{\rm{lower}}}}$ can be approximated by
\begin{equation}\label{eq:P_lower_b}
  {\hat P_{{\rm{lower}}}} \approx Q_1^2Q_2^2\left( {{Q_5}\hat P_{{\rm{lower,1}}}^{\max } + {Q_3} + \hat P_{{\rm{lower,3}}}^{\max }} \right).
\end{equation}
Since $Q_1$, $Q_2$, and $Q_3$ are constants, the maximizing problem for ${\hat P_{{\rm{lower}}}}$ is equivalent to maximizing $f\left( {{b_{\rm{L}}}} \right)$, which is defined by
\begin{equation}\label{eq:f_b_L}
  f\left( {{b_{\rm{L}}}} \right){\rm{ = }}{Q_5}\hat P_{{\rm{lower,1}}}^{\max } + \hat P_{{\rm{lower,3}}}^{\max }{\rm{ = }}{Q_5}\left( {{{\kappa_{\rm B}\kappa_{\rm M}}} - {Q_6}{2^{\frac{{ - \left( {b - {b_{\rm{L}}}} \right)}}{{{L_{\rm{B}}} + {L_{\rm{M}}} - 4}}}}} \right) + 2{Q_4}\cos \left( {\frac{\pi }{{\sqrt {{2^{{b_{\rm{L}}} - {\rm{2}}}}} }}} \right).
\end{equation}
Calculating the derivative of $f\left( {{b_{\rm{L}}}} \right)$ and further setting it to zero, we get
\begin{equation}\label{eq:D_f_b_L}
  \frac{{{Q_5}{Q_6}\ln 2}}{{{L_{\rm{B}}} + {L_{\rm{M}}} - 4}}{2^{\frac{{{b_{\rm{L}}} - b}}{{{L_{\rm{B}}} + {L_{\rm{M}}} - 4}}}} = \frac{{2{Q_4}\pi \ln 2}}{2}{2^{\frac{{{\rm{2}} - {b_{\rm{L}}}}}{2}}}\sin \left( {\frac{\pi }{{\sqrt {{2^{{b_{\rm{L}}} - {\rm{2}}}}} }}} \right).
\end{equation}
Apparently, this equation is intractable due to the coupling of exponential and trigonometric functions. To obtain closed-form result of \eqref{eq:D_f_b_L} and insights about the bit partitioning strategy, we make an assumption that the $b_{\rm L}$ is large enough so that $\sin \frac{\pi }{{\sqrt {{2^{{b_{\rm{L}}} - {\rm{2}}}}} }} \approx \frac{\pi }{{\sqrt {{2^{{b_{\rm{L}}} - {\rm{2}}}}} }}$ is satisfied. Under this assumption, the bit assignment can be obtained as
\begin{equation}\label{eq:b_L_opt}
  b_{\rm{L}}^{\dag} \approx  \frac{b}{{{L_{\rm{B}}} + {L_{\rm{M}}} - 3}} + \frac{{{L_{\rm{B}}} + {L_{\rm{M}}} - 4}}{{{L_{\rm{B}}} + {L_{\rm{M}}} - 3}}{\log _2}\frac{{4\left( {{L_{\rm{B}}} + {L_{\rm{M}}} - 4} \right)\left( {{N_{\rm{R}}} - 1} \right)K_{\rm{B}}^2K_{\rm{M}}^2{\pi ^2}}}{{\left( {{K_{\rm{B}}} + 1} \right)\left( {{K_{\rm{M}}} + 1} \right){Q_6}}}.
\end{equation}
The feedback bits allocated to $b_{\rm L}^{\dag}$ are not only dependent on environment parameters $ {L_{\rm{B}}}$, ${L_{\rm{M}}}$, ${K_{\rm{B}}}$, and ${K_{\rm{M}}}$ but also increase with the RIS's system parameter $N_{\rm R}$. This makes intuitive sense because under the optimal reflection phase design of RIS, the expected received signal power is proportional to $N_{\rm R}^2$ and $N_{\rm R}$ when the signal propagates through the cascaded LoS path and cascaded NLoS paths, respectively \cite{SAoS}. From this perspective, the growth of $N_{\rm R}$ brings more obvious received energy improvement for the cascaded LoS path, and thus more bits should be allocated to $b_{\rm L}$ to reduce the quantization error. Under the integer constraint, we have
\begin{equation}\label{eq:b_L_int}
  b_{{\rm{L}}}^{{\rm{\dag,I}}} = \left\{ \begin{aligned}
&\left\lceil {b_{{\rm{L}}}^{\rm{\dag}}} \right\rceil ,{\rm{ if }\;}f\left( {\left\lceil {b_{{\rm{L}}}^{\rm{\dag}}} \right\rceil } \right) \ge f\left( {\left\lfloor {b_{{\rm{L}}}^{\rm{\dag}}} \right\rfloor } \right)\\
&\left\lfloor {b_{{\rm{L}}}^{\rm{\dag}}} \right\rfloor ,{\rm{ otherwise}}
\end{aligned} \right..
\end{equation}
Following this, the bit partitioning strategy between $b_{\rm L}$ and $b_{\rm N}$ will be given by
\begin{equation}\label{eq:b4LoS_NLoS}
  \left\{ \begin{aligned}
&b_{{\rm{L}}}^\#  = \min \left\{ {\max \left\{ {0,b_{{\rm{L}}}^{{\rm{\dag,I}}} } \right\},{b}} \right\}\\
&b_{{\rm{N}}}^\#  = {b} - b_{{\rm{L}}}^\#
\end{aligned} \right..
\end{equation}\par
In general, we derived closed-form expressions to divide $b$ into four parts and generate our proposed cascaded codebook. The adaptive bit partitioning strategy starts by splitting $b$ into $b_{\rm{L}}^\# $ and $b_{\rm{N}}^\# $ according to \eqref{eq:b4LoS_NLoS}, and then $b_{\rm{L}}^\# $ and $b_{\rm{N}}^\# $ are further separated as $\{b_{{\rm{B,L}}}^\#,b_{{\rm{M,L}}}^\# \}$ and $\{b_{{\rm{B,N}}}^\#,b_{{\rm{M,N}}}^\# \}$ according to \eqref{eq:b4LoS} and \eqref{eq:b4NLoS}, respectively. Note that we do not claim optimality of our bit partitioning strategy due to the approximation and assumption that we  made for the purpose of deriving analytical solutions and useful insights.

%%%%%%%%%%%%%%%%%%%%%%%%%%%%%%%%%%%%%%%%%%%%%%%%%%%%%%%%%%%%%%%%%%%%%%%%%%%%%%%%%%%%%%%%%%%%%%

%%%%%%%%%%%%%%%%%%%%%%%%%%%%%%%%%%%%%%%%%%%%%%%%%%%%%%%%%%%%%%%%%%%%%%%%%%%%%%%%%%%%%%%%%%%%%%
\section{Numerical Results}\label{sec:6}
In this section, numerical results are presented to demonstrate the gain in ergodic rate by using the proposed limited feedback with cascaded codebook over the scheme extended from \cite{Naive_RVQ}, where ${\bf g}_{\rm B}$ and ${\bf g}_{\rm M}$ are quantized by the same naive RVQ codebook. Through simulations, the efficiency of the adaptive bit partitioning is also verified. We consider the BS is equipped with a $2\times 4$ UPA whose antenna spacing is half wavelength. The elements' distance in RIS is also set as half wavelength. In the Cartesian coordinates, the BS, RIS, and MS are deployed in ${\bf p}_{\rm B}=(100,-100,10)$, ${\bf p}_{\rm R}=(0,0,0)$, and ${\bf p}_{\rm M}=(4,5,-3)$, respectively. The AoAs and AoDs of the LoS path can be calculated from the relative locations of the BS, RIS, and MS. As for the NLoS paths, we assume $\{\theta_l\}_{l=2}^{L_{\rm B}}$, $\{\theta_{l,{\rm r}}\}_{l=2}^{L_{\rm B}}$, and $\{\theta_{i,{\rm t}}\}_{i=2}^{L_{\rm M}}$ are randomly distributed in ${\mathcal U}[0, \pi]$, and $\{\phi_l\}_{l=2}^{L_{\rm B}}$, $\{\phi_{l,{\rm r}}\}_{l=2}^{L_{\rm B}}$, and $\{\phi_{i,{\rm t}}\}_{i=2}^{L_{\rm M}}$ are randomly distributed in ${\mathcal U}[-\frac{\pi}{2}, \frac{\pi}{2}]$. The noise power ${\sigma^2}$ is set as $1$.
\subsection{Tightness of the Upper Bound on Ergodic Rate Loss}
\begin{figure}[!t]
\centering
    \includegraphics[width=0.47\textwidth]{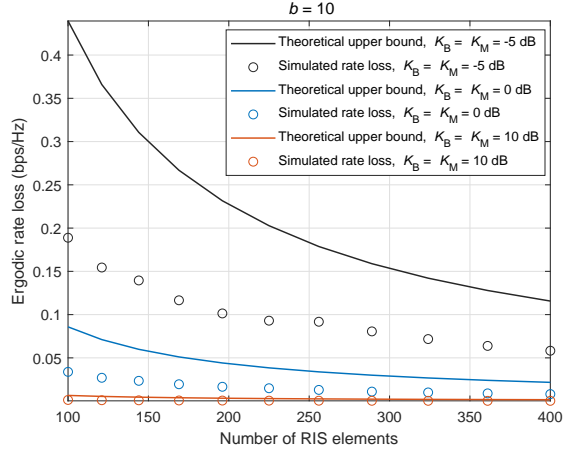}
\caption{Theoretical upper bound and Monte Carlo result of ergodic rate loss as a function of $N_{\rm R}$ for $E=0$ dB, $b=10$, $L_{\rm B}=3$, $L_{\rm M}=10$, $(K_{\rm B}, K_{\rm M})\in \{(-5,-5),(0,0),(10,10)\}$.}
\label{Fig.A1}
\vspace{-0.5cm}
\end{figure}

In Fig. \ref{Fig.A1}, we compare the theoretical upper bound and Monte Carlo result of ergodic rate loss \eqref{eq:rate_gap} in scenarios with different Rician factors. Since we analyze the ergodic rate in large array regime, it is seen that the upper bound gradually presses close to the simulated result when the size of RIS increases. Furthermore, under the strong LoS scenario where the Rician factor is large, the gap between upper bound and simulated result is significantly reduced. This result coincides with our analysis because \emph{Lemma \ref{Lemma:2}} is derived for scenarios with sufficient large Rician factors.
\par
\subsection{Efficiency of Bit Partitioning}
\begin{figure}[!t]
\centering
    \includegraphics[width=0.47\textwidth]{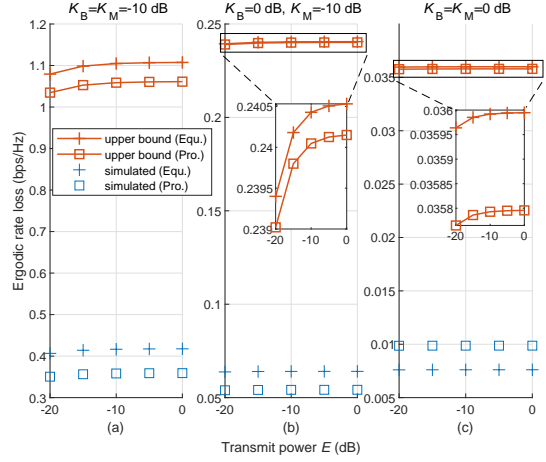}
\caption{Theoretical upper bound and Monte Carlo result of ergodic rate loss as a function of transmit power under different bit allocation schemes (our proposal (Pro.); equally partitioned scheme (Equ.)) for $b=20$, $N_{\rm R}=12\times 20$, $L_{\rm B}=3$, $L_{\rm M}=17$, $(K_{\rm B}, K_{\rm M})\in \{(-10,-10),(0,-10),(0,0)\}$.}
\label{Fig.A2}
\vspace{-0.5cm}
\end{figure}

To verify the efficiency of our proposed bit partitioning strategy, in Fig. \ref{Fig.A2} we compare the ergodic rate with equally partitioned scheme. In Fig. \ref{Fig.A2}(a) where $K_{\rm B}=K_{\rm M}=-10$ dB, the gaps between our proposal and equally partitioned scheme show that the rate loss can be reduced in both upper bound and simulation when feedback bits are properly divided, which verifies the efficiency of our adaptive bit partitioning strategy. However, the upper bound of two schemes gets closer and closer when Rician factors keep growing, as we can see from Figs. \ref{Fig.A2}(b) and (c). We know that the LoS path will gradually dominate the channel, and the NLoS paths will become more and more negligible as the Rician factor increases. Thus, most feedback bits will assigned to $b_{\rm L}$, which is further equally divided. Recall that quantization for the LoS path is to align phases of the LoS and NLoS components. Now that NLoS component is insignificant, feedback bits for LoS path lose their mission. That is to say, $10$ or $5$ bits assigned to $b_{\rm B,L}$ and $b_{\rm M,L}$ produce the same effect. As a matter of fact, equal bit allocation is a special case of the proposed adaptive bit partitioning strategy when Rician factors are large. In Fig. \ref{Fig.A2}(c) where $K_{\rm B}=K_{\rm M}=0$ dB, we also notice that although the proposed strategy has a lower theoretical upper bound on ergodic rate loss than the equal allocation scheme, the Monte Carlo results of rate loss, however, is opposite. This can be explained by the gap between upper bound and simulated result. Utilizing our proposal, the upper bound on rate loss is depressed. However, the gap between actual rate loss and upper bound is larger than the profit we obtained. Besides, in spite of the tininess, the NLoS paths still have energy. Given that the equal allocation scheme assigns more bits to the NLoS component than to our proposed bit partitioning strategy in the strong LoS case, it can gain little performance improvement. However, this improvement is so tiny that we can still regard the equal bit allocation scheme as a special case of our proposal.
\par
\subsection{Equal Path Number}
In this subsection, we show the ergodic rate and bit partitioning results versus Rician factors when paths in segmented channels are equal. To demonstrate the effects of Rician factors, we increase $K_{\rm B}/K_{\rm M}$ with three fixed $K_{\rm M}$. The analytical (Ana.) upper bound and Monte Carlo (M.C.) result of ergodic rate utilizing perfect (Per.) CSI are plotted as baselines. Note that in the figures, quantized (Qua.) and naive (Nav.) quantized CSI represent the CSI obtained by our proposal and scheme extended from \cite{Naive_RVQ}, respectively.\par

\begin{figure}[!t]
\centering
\subfigure[]{
\label{Fig.B1} %% label for first subfigure
\includegraphics[width=0.47\textwidth]{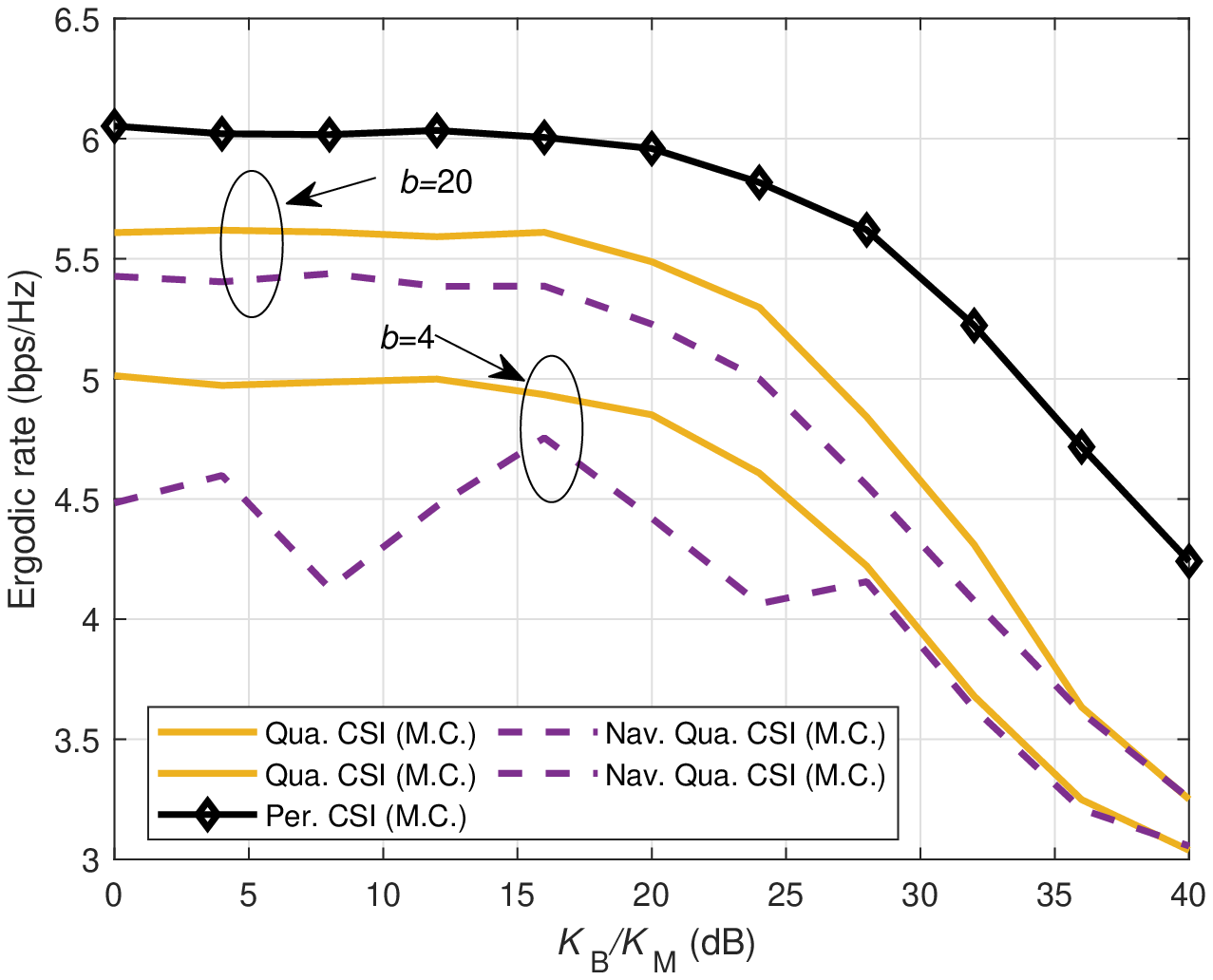}}
\subfigure[]{
\label{Fig.B2} %% label for second subfigure
\includegraphics[width=0.47\textwidth]{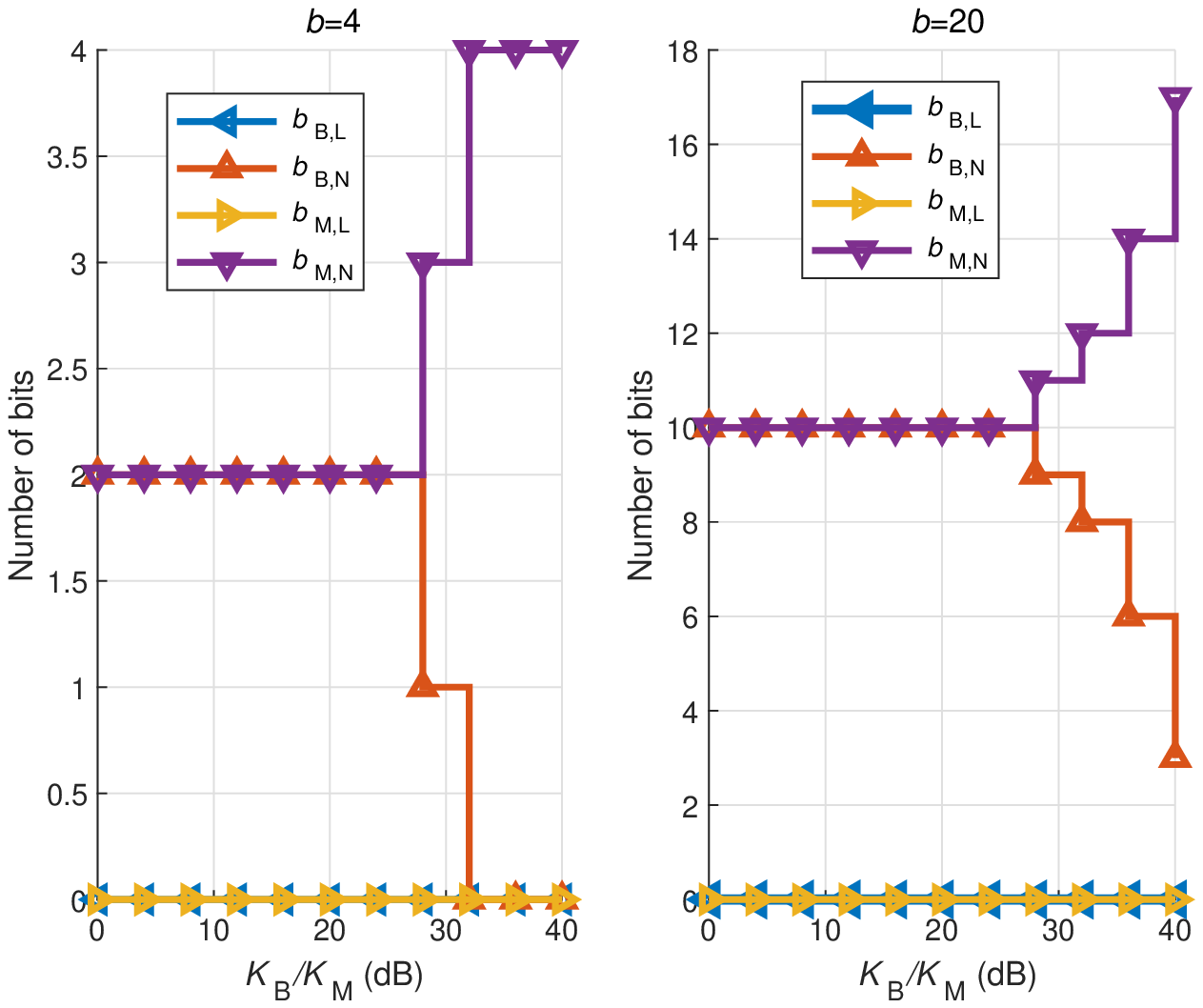}}
\vspace{-0.5cm}
\caption{Ergodic rate and bit partitioning results versus $K_{\rm B}$ when path numbers are equal for $K_{\rm M}=-30$ dB, $E=0$ dB, $L_{\rm B}=L_{\rm M}=6$, $N_{\rm R}=3\times 5$.}
\label{Fig.B1B2} %% label for entire figure
\vspace{-0.5cm}
\end{figure}
In Fig. \ref{Fig.B1B2}, $K_{\rm M}$ is set as $-30$ dB, which means the RIS--MS channel is dominated by the NLoS component. Fig. \ref{Fig.B1} reveals that the gap between analytical upper bound and Monte Carlo result of ergodic rate is large and further enlarged with increasing $K_{\rm B}$. In comparison with naive RVQ quantization, the cascaded codebook scheme achieves obvious rate gain. The feedback bit partitioning results with $b=4$ and $b=20$ are plotted in Fig. \ref{Fig.B2}. As $K_{\rm M}\to 0$ results in weak cascaded LoS path, we observe that no feedback bits are allocated to $b_{\rm B,L}$ and $b_{\rm M,L}$. Furthermore, when both $K_{\rm B}$ and $K_{\rm M}$ are small, we have ${\log _2}\frac{\kappa_{\rm M}}{\kappa_{\rm B}}\to 0$. Hence, \eqref{eq:b_B,N_opt_1} is reduced to equal bit allocation as also shown in Fig. \ref{Fig.B2}. When $K_{\rm B}$ continues to increase, $b_{\rm B,N}$ gradually decrease, which is consistent with the analytical results for case 1 in Section \ref{sec:5.2}.

\begin{figure}[!t]
\centering
\subfigure[]{
\label{Fig.B3} %% label for first subfigure
\includegraphics[width=0.47\textwidth]{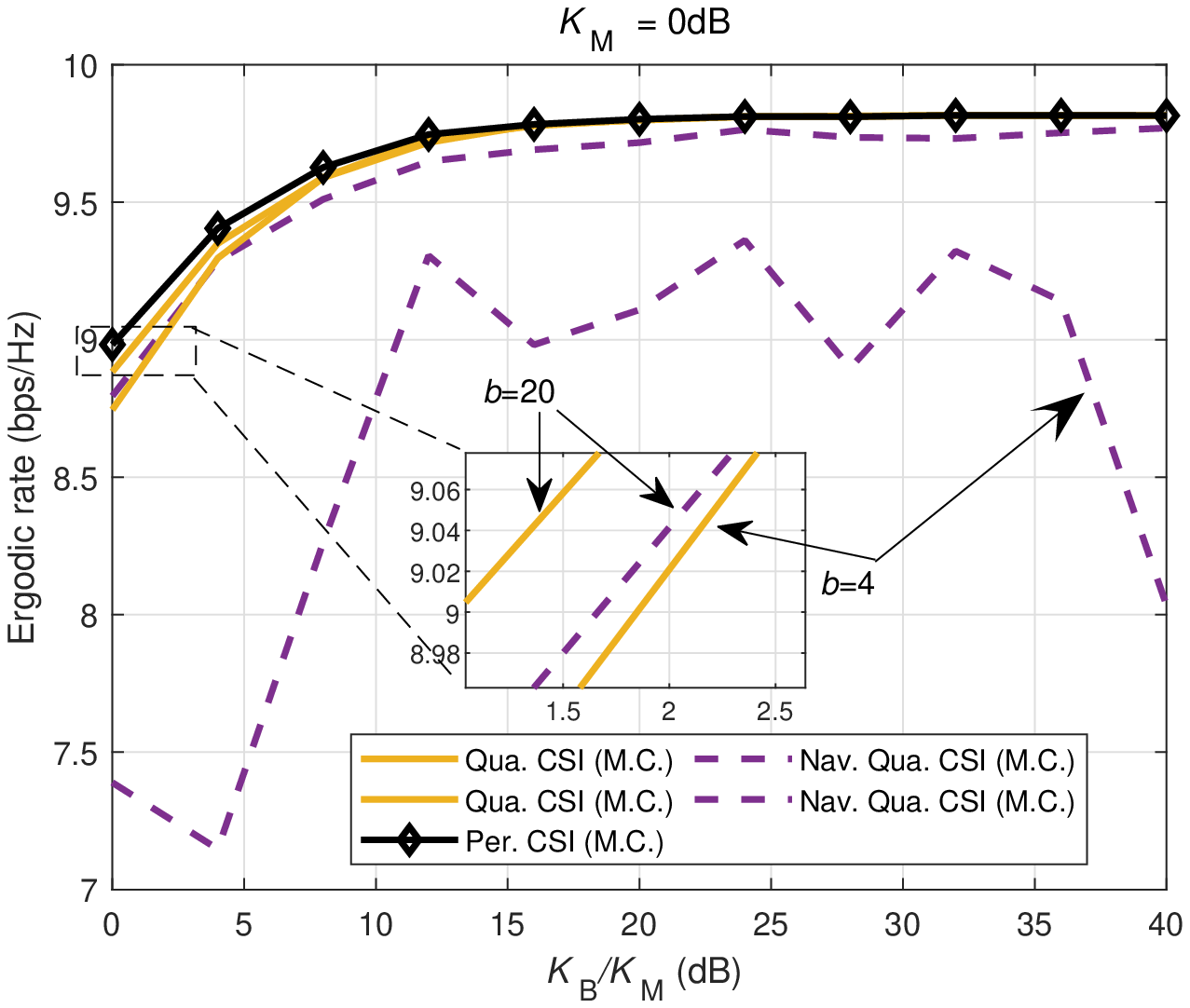}}
\subfigure[]{
\label{Fig.B4} %% label for second subfigure
\includegraphics[width=0.47\textwidth]{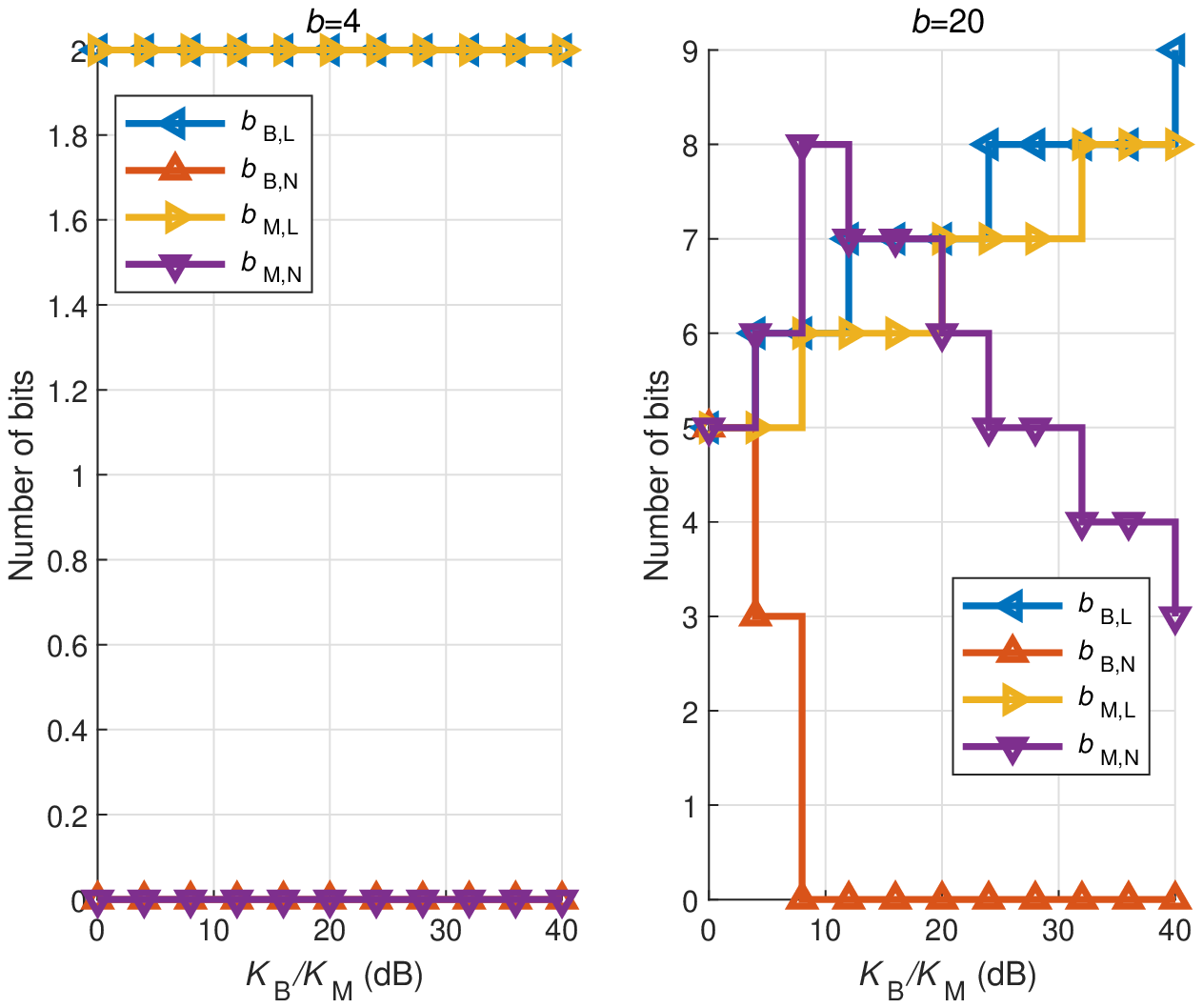}}
\vspace{-0.5cm}
\caption{Ergodic rate and bit partitioning results versus $K_{\rm B}$ when path numbers are equal for $K_{\rm M}=0$ dB, $E=0$ dB, $L_{\rm B}=L_{\rm M}=6$, $N_{\rm R}=3\times 5$.}
\label{Fig.B3B4} %% label for entire figure
\vspace{-0.5cm}
\end{figure}
In Fig. \ref{Fig.B3B4}, we consider the scenario where the LoS and NLoS components of the RIS--MS channel are even by setting $K_{\rm M}=0$ dB. It is observed from \ref{Fig.B3} that the gap between theoretical upper bound and simulation of ergodic rate is narrowed when the cascaded LoS is no longer negligible. The ergodic rate with CSI quantized by our proposed cascaded codebook is nearly identical to that with perfect CSI. Compared with naive RVQ codebook utilizing $b=20$ bits, we can achieve comparable performance with only $b=4$ bits. The bit partitioning results in Fig. \ref{Fig.B4} show that when total bits are inadequate to guarantee quantization accuracy of the primary path gains, all bits are equally assigned to $b_{\rm B,L}$ and $b_{\rm M,L}$. For $b=20$ and $K_{\rm B}=K_{\rm M}=0$ dB, it is observed that $b_{\rm B,L}=b_{\rm B,N}=b_{\rm M,L}=b_{\rm M,N}=5$. This outcome makes intuitive sense because path gains in the BS--RIS and RIS--MS channels are symmetric. The augmentation of $K_{\rm B}$ results in two variation trends. On the one hand, due to the domination of LoS paths, $b_{\rm B,L}$ and $b_{\rm M,L}$ increase consecutively. On the other hand, although $b_{\rm N}$ goes down, $b_{\rm M,N}$ has a transient ascent at the beginning. It is coincident with our analysis that when ${\tilde{\bf{ g}}_{{\rm{B,N}}}}$ becomes less important, more bits should be allocated to quantize ${\tilde{\bf{ g}}_{{\rm{M,N}}}}$.

\begin{figure}[!t]
\centering
\subfigure[]{
\label{Fig.B5} %% label for first subfigure
\includegraphics[width=0.47\textwidth]{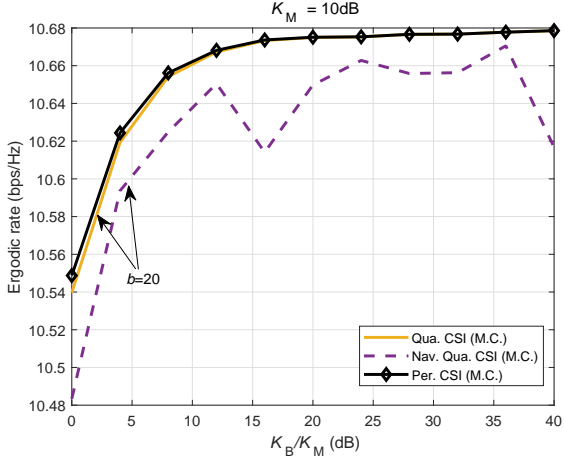}}
\subfigure[]{
\label{Fig.B6} %% label for second subfigure
\includegraphics[width=0.47\textwidth]{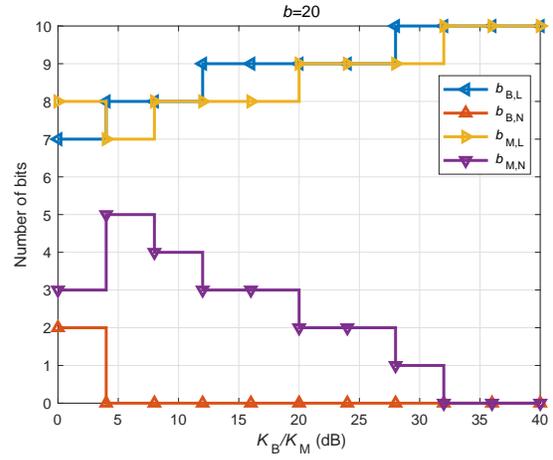}}
\vspace{-0.5cm}
\caption{Ergodic rate and bit partitioning results versus $K_{\rm B}$ when path numbers are equal for $K_{\rm M}=10$ dB, $E=0$ dB, $L_{\rm B}=L_{\rm M}=6$, $N_{\rm R}=3\times 5$.}
\label{Fig.B5B6} %% label for entire figure
\vspace{-0.5cm}
\end{figure}
We further increase $K_{\rm M}$ to $10$ dB to obtain results in a strong LoS scenario, as shown in Fig. \ref{Fig.B5B6}. As feedback bits will all be assigned to $b_{\rm B,L}$ and $b_{\rm M,L}$ when $b$ is insufficient, we only consider $b=20$. In comparison with Figs. \ref{Fig.B1} and \ref{Fig.B3}, we can see from Fig. \ref{Fig.B5} that the entire ergodic rate is promoted as $K_{\rm M}$ increases and our proposal is more robust than the naive RVQ quantization strategy. With LoS paths having more energy than NLoS paths, there is no equal bit allocation appearing in Fig. \ref{Fig.B6}.

\subsection{Equal Rician Factor}

\begin{figure}[!t]
\centering
\subfigure[]{
\label{Fig.C1} %% label for first subfigure
\includegraphics[width=0.47\textwidth]{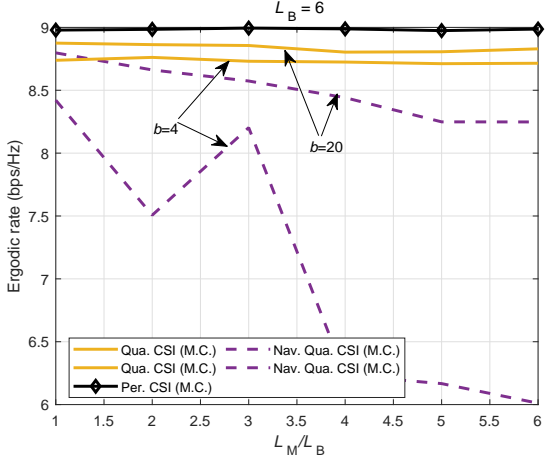}}
\subfigure[]{
\label{Fig.C2} %% label for second subfigure
\includegraphics[width=0.47\textwidth]{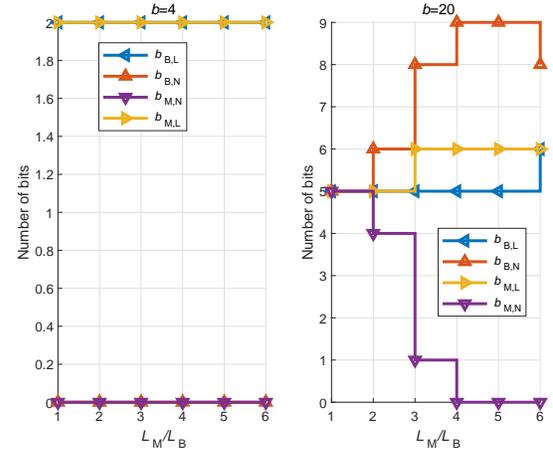}}
\vspace{-0.5cm}
\caption{Ergodic rate and bit partitioning results versus $L_{\rm M}$ when Rician factors are equal for $L_{\rm B}=6$, $E=0$ dB, $K_{\rm B}=K_{\rm M}=0$ dB, $N_{\rm R}=3\times 5$.}
\label{Fig.C1C2} %% label for entire figure
\vspace{-0.5cm}
\end{figure}
In Fig. \ref{Fig.C1C2}, the ergodic rate and bit partitioning results are presented in the case where $K_{\rm B}=K_{\rm M}=0$ dB. We set $L_{\rm B}$ as a reference and increase $L_{\rm M}/L_{\rm B}$. Fig. \ref{Fig.C1} illustrates that the proposed cascaded codebook performs fairly stable than the naive RVQ codebook when the number of paths changes. Furthermore, the ergodic rate goes up steadily when more feedback bits are available. Similar to Fig. \ref{Fig.B4}, all feedback bits are equally allocated to quantize LoS paths when $b$ is severely limited. For $b=20$ and $L_{\rm M}/L_{\rm B}=1$, the adaptive bit partitioning strategy is degraded into an equal bit allocation. Given that the number of paths does not alter the energy between LoS component and NLoS component, we see that the increasing $L_{\rm M}$ only slightly affects $b_{\rm L}$ and $b_{\rm N}$. In Fig. \ref{Fig.C2}, the process of bit segmentation mainly occurs between $b_{\rm B,N}$ and $b_{\rm M,N}$. When paths in the RIS--MS channel are richer, more bits will be allocated to $b_{\rm B,N}$ because each NLoS path in the BS--RIS has a larger energy.

%\begin{figure}[!t]
%\centering
%    \includegraphics[width=0.5\textwidth]{C1.eps}
%\caption{?????.}
%\label{Fig.C1}
%\end{figure}\par
%
%\begin{figure}[!t]
%\centering
%    \includegraphics[width=0.5\textwidth]{C2.eps}
%\caption{?????.}
%\label{Fig.C2}
%\end{figure}\par

\subsection{Increased Feedback Bits}
\begin{figure}[!t]
\centering
\subfigure[]{
\label{Fig.D1} %% label for first subfigure
\includegraphics[width=0.47\textwidth]{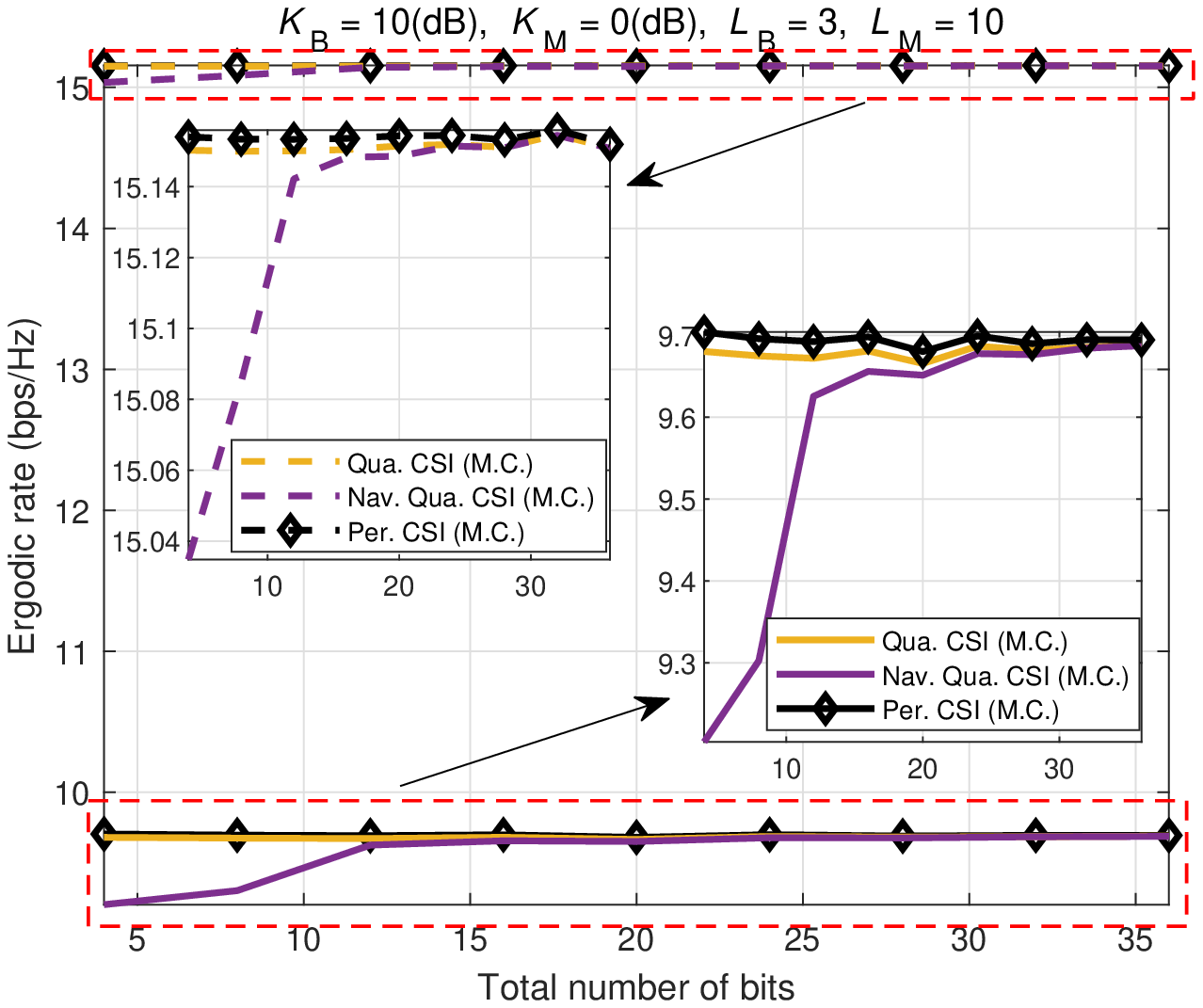}}
\subfigure[]{
\label{Fig.D2} %% label for second subfigure
\includegraphics[width=0.47\textwidth]{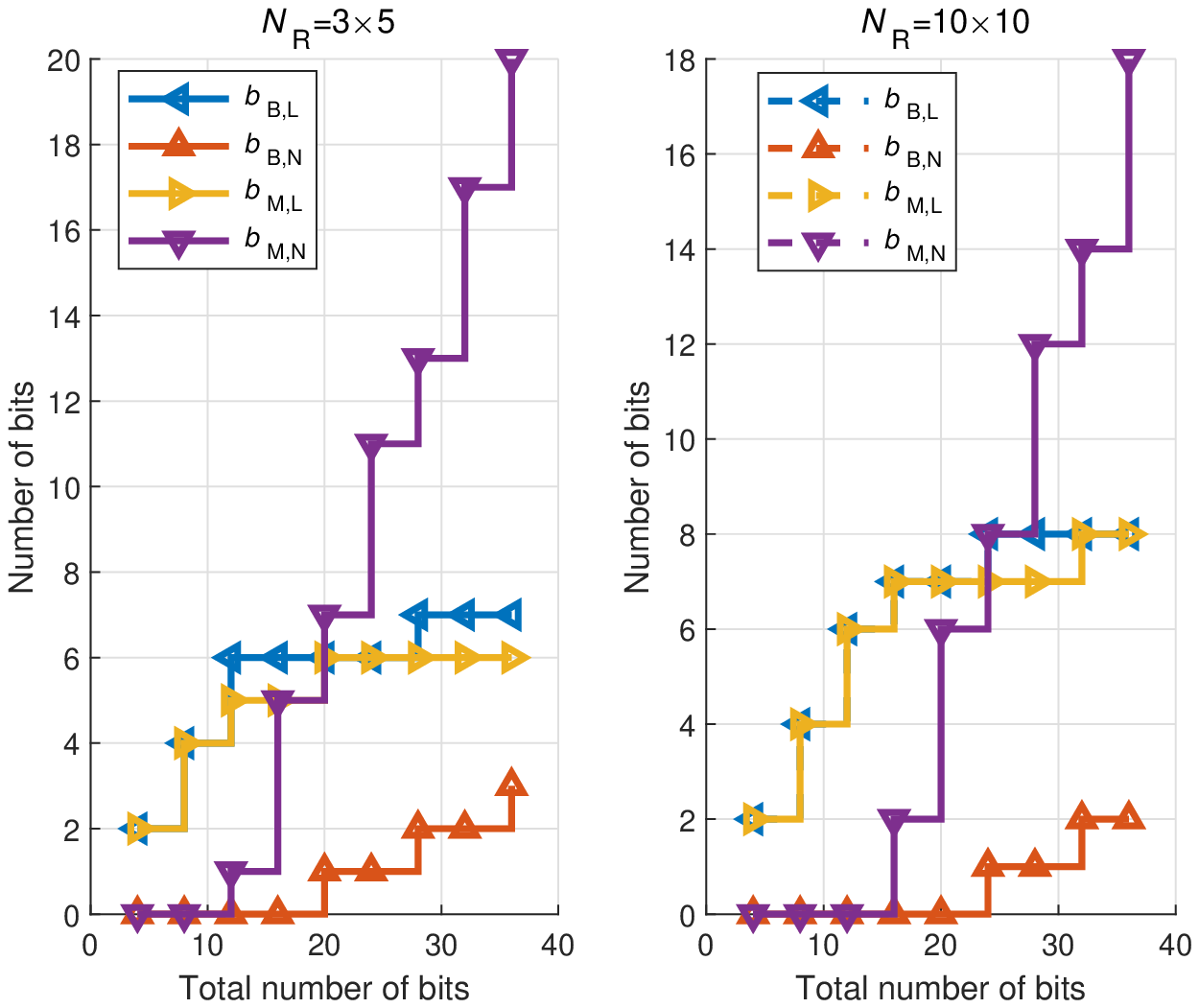}}
\vspace{-0.5cm}
\caption{Ergodic rate and bit partitioning results versus $b$ when segmented channels are asymmetric for $N_{\rm R}=3\times 5$ (solid lines) and $N_{\rm R}=10\times 10$ (dash lines).}
\label{Fig.D1D2} %% label for entire figure
\vspace{-0.5cm}
\end{figure}
Fig. \ref{Fig.D1D2} shows the effects of $b$ and $N_{\rm R}$ on the ergodic rate and bit partitioning. We can observe from Fig. \ref{Fig.D1} that both the cascaded codebook and the naive RVQ codebook can tightly approach the performance of perfect CSI when feedback bits are abundant. However, in the low feedback bit regime, even utilizing $4$ bits, our proposal can still perform reasonably close to the ideal case. This is particularly crucial for quantizing the frequently changing path gains with limited feedback bits. Fig. \ref{Fig.D1} likewise shows that as the size of RIS is enlarged, the ergodic rate increases while the rate loss for the same $b$ decreases. The bit partitioning results described in Fig. \ref{Fig.D2} show that $b_{\rm B,L}$ and $b_{\rm M,L}$ will be promoted first when more feedback bits are available because LoS paths are dominated for $K_{\rm B}=10$ dB and $K_{\rm M}=0$ dB. After LoS paths yield less and less margin, more extra bits will be allocated to quantize NLoS paths. It is interesting to see that $b_{\rm M,N}$ increases dramatically than $b_{\rm B,N}$, the outcome is consistent with the discussion in Section \ref{sec:5.2} that greater $K_{\rm B}$ and $L_{\rm M}$ render more allocation of bits to $b_{\rm M,N}$. Finally, in the high $b$ regime, $b_{\rm L}$ is larger for $N_{\rm R}=10\times 10$ than for $N_{\rm R}=3\times 5$. The reason is that the growth of $N_{\rm R}$ brings more obvious received energy improvement for the cascaded LoS path.

%\begin{figure}[!t]
%\centering
%    \includegraphics[width=0.5\textwidth]{D1.eps}
%\caption{Ergodic rate as a function of total feedback bits under different quantization schemes (Per. CSI(Ana.) : analytical upper bound with perfect CSI; Per. CSI(M.C.) : Monte Carlo result with perfect CSI; Qua. CSI (M.C.) : Monte Carlo result of our quantized CSI; Nav. Qua. CSI(M.C.): Monte Carlo result of naive CSI quantization scheme extended from \cite{Naive_RVQ}) for $N_{\rm R}=3\times 5$ (solid lines) and $N_{\rm R}=10\times 10$ (dash lines).}
%\label{Fig.D1}
%\end{figure}\par
%\begin{figure}[!t]
%\centering
%    \includegraphics[width=0.5\textwidth]{D2.eps}
%\caption{\textcolor[rgb]{1,0,0}{To be continued.}Theoretical upper bound and Monte Carlo result of ergodic rate loss as a function of transmit SNR under different bit allocation scheme (Par. : Partitioned with our proposal; Equ. : Equally partitioned) for $b=20$, $N_{\rm R,v}=12$, $N_{\rm R,v}=20$, $L_{\rm B}=3$, $L_{\rm M}=17$, $(K_{\rm B}, K_{\rm M})\in \{(-10,-10),(0,-10),(0,0)\}$.}
%\label{Fig.D2}
%\end{figure}\par

\section{Conclusions}\label{sec:7}
In this study, we investigated the channel feedback in an RIS-assisted FDD system where the BS and MS build a reliable communication link through RIS. Since RIS segments the BS-MS channel into two high-dimensional sub-channels, conventional feedback schemes are inapplicable due to the unbearable overhead. To address this, we parameterized the sub-channels and focused on the feedback of the channel path gains because path directions can be obtained at the BS through the uplink. A cascaded codebook was proposed for path gain quantization. Different from existing works, the proposed codeword was synthesized by four parts that can respectively quantize the LoS and NLoS path gains in two sub-channels. We further investigated the performance of the cascaded codebook by deriving an upper bound on ergodic rate loss caused by quantization error. It was found that the rate loss can be cut down by properly allocating the feedback bits used to generate the cascaded codebook. On the basis of this observation, an adaptive bit partitioning strategy was proposed to minimize the upper bound on ergodic rate loss. Numerical results verified the efficiency of our bit allocation algorithm and showed its adaptivity to different environment and system parameters.

\appendices
\section{}\label{App:A}
In this Appendix, we analyze ${{\bf{A}}^H}{\bf{A}}$ in the large-dimension regime where ${N_{\rm{B}}}\rightarrow\infty$ and ${N_{\rm{R}}}\rightarrow\infty$. According to the definition of the cascaded path direction matrix, we have
\begin{equation}\label{eq:A-1}
{{\bf{A}}^H}{\bf{A}}{\rm{ = }}\left[ {\begin{array}{*{20}{c}}
{{\bf{a}}_{{\rm{E}},1}^H{{\bf{a}}_{{\rm{E}},1}}}& \cdots &{{\bf{a}}_{{\rm{E}},1}^H{{\bf{a}}_{{\rm{E}},{L_{\rm{M}}}{L_{\rm{B}}}}}}\\
 \vdots & \ddots & \vdots \\
{{\bf{a}}_{{\rm{E}},{L_{\rm{M}}}{L_{\rm{B}}}}^H{{\bf{a}}_{{\rm{E}},1}}}& \cdots &{{\bf{a}}_{{\rm{E}},{L_{\rm{M}}}{L_{\rm{B}}}}^H{{\bf{a}}_{{\rm{E}},{L_{\rm{M}}}{L_{\rm{B}}}}}}
\end{array}} \right].
\end{equation}
The element in the $n_1$-th column and the $n_2$-th row is taken as an example. For clear and concise description, we denote indices ${l_1} \buildrel \Delta \over = {v_{{L_{\rm{M}}}}}( {{n_1}} )$, ${l_2} \buildrel \Delta \over = {v_{{L_{\rm{M}}}}}( {{n_2}} )$, ${i_1} \buildrel \Delta \over = {h_{{L_{\rm{M}}}}}( {{n_1}} )$, and ${i_2} \buildrel \Delta \over = {h_{{L_{\rm{M}}}}}( {{n_2}} )$; and array response vectors ${{\bf{a}}_{{\rm{R,t,}}i}} \buildrel \Delta \over = {{\bf{a}}_{\rm{R}}}( {{\Theta _{i{\rm{,t}}}},{\Phi _{i{\rm{,t}}}}} )$, ${{\bf{a}}_{{\rm{R,r,}}l}} \buildrel \Delta \over = {{\bf{a}}_{\rm{R}}}( {{\Theta _{l{\rm{,r}}}},{\Phi _{l{\rm{,r}}}}} )$, and ${{\bf{a}}_{{\rm{B,}}l}} \buildrel \Delta \over = {{\bf{a}}_{\rm{B}}}( {{\Theta _l},{\Phi _l}} )$. Afterward, we have
\begin{equation}\label{eq:A-2}
  \begin{aligned}
&{\bf{a}}_{{\rm{E}},{n_1}}^H{{\bf{a}}_{{\rm{E}},{n_2}}}= {{\bm{\psi }}^T}{\rm{diag}}\left( {{\bf{a}}_{{\rm{R,t}},{i_1}}^H} \right){{\bf{a}}_{{\rm{R,r,}}{l_1}}}{\bf{a}}_{{\rm{B,}}{l_1}}^H{{\bf{a}}_{{\rm{B,}}{l_2}}}{\bf{a}}_{{\rm{R,r,}}{l_2}}^H{\rm{diag}}\left( {{{\bf{a}}_{{\rm{R,t}},{i_2}}}} \right){{\bm{\psi }}^*}\\
&\mathop  = \limits^{\left( a \right)} \frac{1}{{{N_{\rm{B}}}}}{\alpha _{{l_1},{l_2}}}{e^{j{\beta _{{l_1},{l_2}}}}}{{\bm{\psi }}^T}{\rm{diag}}\left( {{\bf{a}}_{{\rm{R,t}},{i_1}}^H} \right){{\bf{a}}_{{\rm{R,r,}}{l_1}}}{\bf{a}}_{{\rm{R,r,}}{l_2}}^H{\rm{diag}}\left( {{{\bf{a}}_{{\rm{R,t}},{i_2}}}} \right){{\bm{\psi }}^*}\\
&\mathop  = \limits^{\left( b \right)} \frac{1}{{{N_{\rm{B}}}}}{\alpha _{{l_1},{l_2}}}{e^{j{\beta _{{l_1},{l_2}}}}}\left(\frac{1}{{{N_{\rm{R}}}}}\sum\limits_{s = 1}^{{N_{\rm{R}}}} {{e^{j{\Omega _{{l_1},{i_1},s}}}}} \right)\left(\frac{1}{{{N_{\rm{R}}}}}\sum\limits_{c = 1}^{{N_{\rm{R}}}} {{e^{ - j{\Omega _{{l_2},{i_2},c}}}}}\right) = \frac{{{\alpha _{{l_1},{l_2}}}{e^{j{\beta _{{l_1},{l_2}}}}}}}{{{N_{\rm{B}}}N_{\rm{R}}^2}}\sum\limits_{s = 1}^{{N_{\rm{R}}}} {\sum\limits_{c = 1}^{{N_{\rm{R}}}} {{e^{j{\Omega _{{l_1},{i_1},s}} - j{\Omega _{{l_2},{i_2},c}}}}} },
\end{aligned}
\end{equation}
where $(a)$ is obtained from \eqref{eq:B-1}, and $(b)$ is proved in Appendix \ref{App:C}. On the basis of \eqref{eq:A-2}, four cases are considered for ${\bf{a}}_{{\rm{E}},{n_1}}^H{{\bf{a}}_{{\rm{E}},{n_2}}}$ in the following:
\begin{enumerate}
  \item ${n_1} = {n_2} = 1$: The aim of the reflection phase design of RIS in \eqref{eq:optimal_phase} is to align the LoS links of the BS--RIS and RIS--MS channels so that ${\Omega _{1,1,c}} - {\Omega _{1,1,s}} = 0$. Thus, it easily obtains
\begin{equation}\label{eq:A-case-1}
  {\bf{a}}_{{\rm{E}},1}^H{{\bf{a}}_{{\rm{E}},1}} = \frac{1}{{N_{\rm{R}}^2}}\sum\limits_{s = 1}^{{N_{\rm{R}}}} {\sum\limits_{c = 1}^{{N_{\rm{R}}}} {{e^{j{\Omega _{1,1,s}} - j{\Omega _{1,1,c}}}}} }  = 1.
\end{equation}
  \item ${n_1} = {n_2} = n \ne 1$: In this case,
\begin{equation}\label{eq:A-case-2}
  \begin{aligned}
{\bf{a}}_{{\rm{E}},n}^H{{\bf{a}}_{{\rm{E}},n}} = \frac{1}{{N_{\rm{R}}^2}}\sum\limits_{s = 1}^{{N_{\rm{R}}}} {{e^{j\left( {{\Omega _{l,i,s}} - {\Omega _{l,i,s}}} \right)}}}  + \frac{1}{{N_{\rm{R}}^2}}\sum\limits_{s = 1}^{{N_{\rm{R}}}} {\sum\limits_{c \ne s}^{{N_{\rm{R}}}} {{e^{j\left( {{\Omega _{l,i,s}} - {\Omega _{l,i,c}}} \right)}}} }  = \frac{1}{{{N_{\rm{R}}}}} + \frac{1}{{N_{\rm{R}}^2}}\sum\limits_{s = 1}^{{N_{\rm{R}}}} {\sum\limits_{c \ne s}^{{N_{\rm{R}}}} {{e^{j\left( {{\Omega _{l,i,s}} - {\Omega _{l,i,c}}} \right)}}} }
\end{aligned}.
\end{equation}
Considering that the reflection phases have been designed, the second term in \eqref{eq:A-case-2} is the average of $N_{\rm R}(N_{\rm R}-1)$ complex units whose phases are random. Thus, $\frac{1}{{N_{\rm{R}}^2}}\sum\nolimits_{s = 1}^{{N_{\rm{R}}}} {\sum\nolimits_{c \ne s}^{{N_{\rm{R}}}} {{e^{j({\Omega _{l,i,s}} - {\Omega _{l,i,c}})}}} }  \approx 0$ holds and ${\bf{a}}_{{\rm{E}},n}^H{{\bf{a}}_{{\rm{E}},n}}=\frac{1}{{{N_{\rm{R}}}}}$.

  \item ${n_1} \ne {n_2}$ $( {{l_1} \ne {l_2}} )$: According to the orthogonality among array response vectors of the BS described in Appendix \ref{App:B}, when ${N_{\rm{B}}} \to \infty $, ${\bf{a}}_{{\rm{B,}}{l_1}}^H{{\bf{a}}_{{\rm{B,}}{l_2}}} = 0$ holds. Thus, we have ${\bf{a}}_{{\rm{E}},{n_1}}^H{{\bf{a}}_{{\rm{E}},{n_2}}} = 0$ in this case.
  \item ${n_1} \ne {n_2}$ $( {{l_1} = {l_2},{i_1} \ne {i_2}} )$: Since ${l_1} = {l_2}$, we have
\begin{equation}\label{eq:A-case-4}
  {\bf{a}}_{{\rm{E}},{n_1}}^H{{\bf{a}}_{{\rm{E}},{n_2}}} = \frac{1}{{N_{\rm{R}}^2}}\sum\limits_{s = 1}^{{N_{\rm{R}}}} {\sum\limits_{c = 1}^{{N_{\rm{R}}}} {{e^{j{\Omega _{{l_1},{i_1},s}} - j{\Omega _{{l_1},{i_2},c}}}}} }.
\end{equation}
Similar to Case 2, phases of the complex units in \eqref{eq:A-case-4} cannot be aligned. Therefore, ${\bf{a}}_{{\rm{E}},{n_1}}^H{{\bf{a}}_{{\rm{E}},{n_2}}} = 0$ also holds.
\end{enumerate}
To sum up, for the large-dimension regime where ${N_{\rm{B}}}\rightarrow\infty$ and ${N_{\rm{R}}}\rightarrow\infty$, \emph{Property \ref{Them:1}} can be proved.

\section{}\label{App:B}
In this Appendix, we prove that in the large-scale UPA, the array response vectors are asymptotically orthogonal to one another. Using the results from \cite{beam_positioning}, the array response vectors of UPA satisfy
\begin{equation}\label{eq:B-1}
  {\bf{a}}_{{\rm{B,}}{l_1}}^H{{\bf{a}}_{{\rm{B,}}{l_2}}} = \frac{1}{{{N_{\rm{B}}}}}{\alpha _{{l_1},{l_2}}}{e^{j{\beta _{{l_1},{l_2}}}}},
\end{equation}
where
\begin{equation}\label{eq:B-2}
  \left\{ \begin{aligned}
&{\alpha _{{l_1},{l_2}}} = \frac{{\sin \left( {\frac{1}{2}\left( {{\Theta _{{l_2}}} - {\Theta _{{l_1}}}} \right){N_{{\rm{B,v}}}}} \right)\sin \left( {\frac{1}{2}\left( {{\Phi _{{l_2}}} - {\Phi _{{l_1}}}} \right){N_{{\rm{B,h}}}}} \right)}}{{\sin \left( {\frac{1}{2}\left( {{\Theta _{{l_2}}} - {\Theta _{{l_1}}}} \right)} \right)\sin \left( {\frac{1}{2}\left( {{\Phi _{{l_2}}} - {\Phi _{{l_1}}}} \right)} \right)}}\\
&{\beta _{{l_1},{l_2}}} = \frac{1}{2}\left[ {\left( {{\Theta _{{l_2}}} - {\Theta _{{l_1}}}} \right)\left( {{N_{{\rm{B,v}}}} - 1} \right) + \left( {{\Phi _{{l_2}}} - {\Phi _{{l_1}}}} \right)\left( {{N_{{\rm{B,h}}}} - 1} \right)} \right]
\end{aligned} \right..
\end{equation}
If ${l_1} = {l_2}$, it is easy to have ${\bf{a}}_{{\rm{B,}}{l_1}}^H{{\bf{a}}_{{\rm{B,}}{l_2}}} = 1$. When ${l_1} \ne {l_2}$, we define a function $f( x ) \buildrel \Delta \over = | {\frac{{\sin ( {Nx} )}}{{N\sin ( x )}}} |$. The null points of $f( x )$ are $\frac{{k\pi }}{N},k =  \pm 1,2,3, \cdots $. For $N \to \infty $, ${x_0} \buildrel \Delta \over = \frac{1}{2}( {{\Theta _{{l_2}}} - {\Theta _{{l_1}}}} )$ is much larger than the first null point (i.e., $\frac{1}{2}( {{\Theta _{{l_2}}} - {\Theta _{{l_1}}}} ) \gg \frac{{\pi }}{N}$). Without loss of generality, assuming $x_0$ lies between the $k$-th $( {1 \ll k \ll N} )$ and $(k+1)$-th null point (i.e., ${x_0} \in [ {\frac{{k\pi }}{N},\frac{{( {k + 1} )\pi }}{N}} ]$), we have
\begin{equation}\label{eq:B-3}
  f\left( {{x_0}} \right) = \left| {\frac{{\sin \left( {N{x_0}} \right)}}{{N\sin \left( {{x_0}} \right)}}} \right| \approx \left| {\frac{{\sin \left( {N{x_0}} \right)}}{{N{x_0}}}} \right| \le \left| {\frac{1}{{N{x_0}}}} \right| < \left| {\frac{1}{{N\frac{{k\pi }}{N}}}} \right| = \left| {\frac{1}{{k\pi }}} \right| \approx 0,
\end{equation}
where the approximation is due to the fact that when $N \gg k$, $0 \le {x_0} \le \frac{{\left( {k + 1} \right)\pi }}{N} \approx 0$ holds to satisfy $\mathop {\lim }\limits_{{x_0} \to 0} \sin \left( {{x_0}} \right) \approx {x_0}$. On the basis of \eqref{eq:B-3}, we can conclude that when ${N_{{\rm{B,v}}}} \to \infty $ or ${N_{{\rm{B,h}}}} \to \infty $, if ${l_1} \ne {l_2}$, then the array response vectors of UPA are asymptotically orthogonal to one another, that is, ${\bf{a}}_{{\rm{B,}}{l_1}}^H{{\bf{a}}_{{\rm{B,}}{l_2}}} \to 0$.

\section{}\label{App:C}
The inner product between the reflection coefficient vector and the cascaded array response vector is calculated as follows:
\begin{equation}\label{eq:C-1}
  \begin{aligned}
{{\bm{\psi }}^T}{\rm{diag}}\left( {{\bf{a}}_{{\rm{R,t}},{i_1}}^H} \right){{\bf{a}}_{{\rm{R,r,}}{l_1}}} &= {{\bm{\psi }}^T}\left( {{\bf{a}}_{{\rm{R,t}},{i_1}}^* \odot {{\bf{a}}_{{\rm{R,r,}}{l_1}}}} \right) = {{\bm{\psi }}^T}\left( {\left( {{\bf{a}}_{{\rm{R,t,v}},{i_1}}^* \odot {{\bf{a}}_{{\rm{R,r,v,}}{l_1}}}} \right) \otimes \left( {{\bf{a}}_{{\rm{R,t,h}},{i_1}}^* \odot {{\bf{a}}_{{\rm{R,r,h,}}{l_1}}}} \right)} \right)\\
&  = \frac{1}{{{N_{\rm{R}}}}}{{\bf{\psi }}^T}\left( {\left[ {\begin{array}{*{20}{c}}
1\\
 \vdots \\
{{e^{j\left( {{N_{{\rm{R,v}}}} - 1} \right)\left( {{\Theta _{{l_1}{\rm{,r}}}} - {\Theta _{{i_1}{\rm{,t}}}}} \right)}}}
\end{array}} \right] \otimes \left[ {\begin{array}{*{20}{c}}
1\\
 \vdots \\
{{e^{j\left( {{N_{{\rm{R,h}}}} - 1} \right)\left( {{\Phi _{{l_1}{\rm{,r}}}} - {\Phi _{{i_1}{\rm{,t}}}}} \right)}}}
\end{array}} \right]} \right) \\
&= \frac{1}{{{N_{\rm{R}}}}}\sum\limits_{s = 1}^{{N_{\rm{R}}}} {{e^{j{\Omega _{{l_1},{i_1},s}}}}},
\end{aligned}
\end{equation}
where ${\Omega _{{l_1},{i_1},s}} = {\psi _s} + \left( {{v_{{N_{{\rm{R,h}}}}}}\left( s \right) - 1} \right)\left( {{\Theta _{{l_1}{\rm{,r}}}} - {\Theta _{{i_1}{\rm{,t}}}}} \right) + \left( {{h_{{N_{{\rm{R,h}}}}}}\left( s \right) - 1} \right)\left( {{\Phi _{{l_1}{\rm{,r}}}} - {\Phi _{{i_1}{\rm{,t}}}}} \right)$.

\section{}\label{App:lemma-1}
The expectation of $X_2$ can be expanded as
\begin{equation}\label{eq:APP-E_X_2}
  \begin{aligned}
{\mathbb E} \left\{ {{{\left| {{X_2}} \right|}^2}} \right\} &= {\mathbb E}\left\{ {{{\left| {\left( {1 - \frac{1}{{{N_{\rm{R}}}}}} \right)g_{{\rm{B,1}}}^H\sqrt {\frac{{{K_{\rm{B}}}}}{{{K_{\rm{B}}} + 1}}} {w_{{\rm{B,L}},i_{{\rm{B,L}}}^\# }}g_{{\rm{M,1}}}^H\sqrt {\frac{{{K_{\rm{M}}}}}{{{K_{\rm{M}}} + 1}}} {w_{{\rm{M,L}},i_{{\rm{M,L}}}^\# }}} \right|}^2}} \right\}\\
% &= \frac{{{K_{\rm{B}}}}}{{{K_{\rm{B}}} + 1}}\frac{{{K_{\rm{M}}}}}{{{K_{\rm{M}}} + 1}}\frac{{{{\left( {{N_{\rm{R}}} - 1} \right)}^2}}}{{N_{\rm{R}}^2}} {\mathbb E}\left\{ {{{\left| {g_{{\rm{B,1}}}^H{w_{{\rm{B,L}},i_{{\rm{B,L}}}^\# }}g_{{\rm{M,1}}}^H{w_{{\rm{M,L}},i_{{\rm{M,L}}}^\# }}} \right|}^2}} \right\}\\
 &= \frac{{{K_{\rm{B}}}{K_{\rm{M}}}{{\left( {{N_{\rm{R}}} - 1} \right)}^2}}}{N_{\rm{R}}^2{\left( {{K_{\rm{B}}} + 1} \right)\left( {{K_{\rm{M}}} + 1} \right)}}{\mathbb E}\left\{ {{{\left| {{g_{{\rm{B,1}}}}} \right|}^2}{{\left| {{g_{{\rm{M,1}}}}} \right|}^2}} \right\}.
\end{aligned}
\end{equation}
Given the independence of path gains in the BS--RIS and RIS--MS channel, ${\mathbb E}\{ {{{| {{g_{{\rm{B,1}}}}} |}^2}{{| {{g_{{\rm{M,1}}}}} |}^2}} \} = {\mathbb E}\{ {{{| {{g_{{\rm{B,1}}}}} |}^2}} \}{\mathbb E}\{ {{{| {{g_{{\rm{M,1}}}}} |}^2}} \}$. For the LoS path gain in the BS--RIS channel, we have
\begin{equation}\label{eq:APP-E-g}
  {\mathbb E}\left\{ {{{\left| {{g_{{\rm{B,1}}}}} \right|}^2}} \right\} = {K_{\rm{B}}}\left( {{L_{\rm{B}}} - 1} \right){\mathbb E}\left\{ {{{\left| {{g_{{\rm{B}},{\rm{L}}}}} \right|}^2}} \right\} = {K_{\rm{B}}}\left( {{L_{\rm{B}}} - 1} \right).
\end{equation}
Similarly, ${\mathbb E}\{ {{{| {{g_{{\rm{M,1}}}}} |}^2}} \} = {K_{\rm{M}}}( {{L_{\rm{M}}} - 1} )$.

\section{}\label{App:E}
According to the definitions of $X_1$ and $X_2$ and the independence of path gains in the BS--RIS and RIS--MS channels, we have
\begin{equation}\label{eq:APP-E-1}
  \begin{aligned}
{\mathbb E}\left\{ {\left| {{X_1}} \right|\left| {{X_2}} \right|} \right\}
 = \sqrt {\frac{{{K_{\rm{B}}}{K_{\rm{M}}}}}{{\left( {{K_{\rm{B}}} + 1} \right)\left( {{K_{\rm{M}}} + 1} \right)}}} \frac{{\left( {{N_{\rm{R}}} - 1} \right)}}{{N_{\rm{R}}^2}}{\mathbb E}\left\{ {\left| {{\bf{g}}_{\rm{B}}^H{{\bf{w}}_{{\rm{B}},i_{\rm{B}}^\# }}g_{{\rm{B,1}}}^H} \right|} \right\}{\mathbb E}\left\{ {\left| {{\bf{g}}_{\rm{M}}^H{{\bf{w}}_{{\rm{M}},i_{\rm{M}}^\# }}g_{{\rm{M,1}}}^H} \right|} \right\},
\end{aligned}
\end{equation}
In \eqref{eq:APP-E-1}, ${\mathbb E}\{ {| {{\bf{g}}_{\rm{B}}^H{{\bf{w}}_{{\rm{B}},i_{\rm{B}}^\# }}} || {g_{{\rm{B,1}}}^H} |} \}$ can be further unfolded by substituting the cascaded forms of ${\bf g}_{\rm B}$ and ${{{\bf{w}}_{{\rm{B}},i_{\rm{B}}^\# }}}$, that is,
\begin{equation}\label{eq:APP-E-2}
  \begin{aligned}
{\mathbb E}\left\{ {| {{\bf{g}}_{\rm{B}}^H{{\bf{w}}_{{\rm{B}},i_{\rm{B}}^\# }}} |\left| {g_{{\rm{B,1}}}^H} \right|} \right\} = \sqrt {\frac{1}{{{K_{\rm{B}}} + 1}}}{\mathbb E} \left\{ {| {\sqrt {{K_{\rm{B}}}} g_{{\rm{B,1}}}^H{e^{j{\varepsilon _{i_{{\rm{B,L}}}^\# }}}} + {\bf{g}}_{{\rm{B,N}}}^H{{\bf{w}}_{{\rm{B,N}},i_{{\rm{B,L}}}^\# }}} |\left| {g_{{\rm{B,1}}}^H} \right|} \right\}
\end{aligned}.
\end{equation}
To further simplify \eqref{eq:APP-E-2}, we compare the magnitude square of $\sqrt {{K_{\rm{B}}}} g_{{\rm{B,1}}}^H{e^{j{\varepsilon _{i_{{\rm{B,L}}}^\# }}}}$ and ${\bf{g}}_{{\rm{B,N}}}^H{{\bf{w}}_{{\rm{B,N}},i_{{\rm{B,L}}}^\# }}$. First, it easily obtains
\begin{equation}\label{eq:APP-E-3}
  {\mathbb E}\left\{ {{{\left| {\sqrt {{K_{\rm{B}}}} g_{{\rm{B,1}}}^H{e^{j{\varepsilon _{i_{{\rm{B,L}}}^\# }}}}} \right|}^2}} \right\} = K_{\rm{B}}^2\left( {{L_{\rm{B}}} - 1} \right).
\end{equation}
Exploiting the independence of the vector norm and vector direction \cite{Nihar}, we obtain
\begin{equation}\label{eq:APP-E-4}
  \begin{aligned}
{\mathbb E}\left\{ {{{\left| {{\bf{g}}_{{\rm{B,N}}}^H{{\bf{w}}_{{\rm{B,N}},i_{{\rm{B,L}}}^\# }}} \right|}^2}} \right\} = {\mathbb E}\left\{ {{{\left\| {{{\bf{g}}_{\rm{B}}}} \right\|}^2}} \right\}{\mathbb E}\left\{ {{{\left| {\tilde{\bf{ g}}_{{\rm{B,N}}}^H{{\bf{w}}_{{\rm{B,N}},i_{{\rm{B,L}}}^\# }}} \right|}^2}} \right\}= \left( {{L_{\rm{B}}} - 1} \right)\left( {{K_{\rm{B}}} + 1} \right){\mathbb E}\left\{ {{{\left| {\tilde{\bf{ g}}_{{\rm{B,N}}}^H{{\bf{w}}_{{\rm{B,N}},i_{{\rm{B,L}}}^\# }}} \right|}^2}} \right\}
\end{aligned}.
\end{equation}
As ${{{\bf{w}}_{{\rm{B,N}},i_{{\rm{B,L}}}^\# }}}$ is selected from the RVQ codebook, according to Lemma 1 in \cite{Nihar}, we have
\begin{equation}\label{eq:APP-E-5}
  1 - {2^{ - \frac{{{b_{{\rm{B,N}}}}}}{{{L_{\rm{B}}} - 2}}}}<{\mathbb E}\left\{ {{{\left| {\tilde{\bf{ g}}_{{\rm{B,N}}}^H{{\bf{w}}_{{\rm{B,N}},i_{{\rm{B,L}}}^\# }}} \right|}^2}} \right\} \leq 1.
\end{equation}
On the basis of \eqref{eq:APP-E-3}--\eqref{eq:APP-E-5}, it can be concluded that
\begin{equation}\label{eq:APP-E-6}
  \frac{{\mathbb E}{\left\{ {{{\left| {\sqrt {{K_{\rm{B}}}} g_{{\rm{B,1}}}^H{e^{j{\varepsilon _{i_{{\rm{B,L}}}^\# }}}}} \right|}^2}} \right\}}}{{\mathbb E}{\left\{ {{{\left| {{\bf{g}}_{{\rm{B,N}}}^H{{\bf{w}}_{{\rm{B,N}},i_{{\rm{B,L}}}^\# }}} \right|}^2}} \right\}}} \geq \frac{{K_{\rm{B}}^2}}{{{K_{\rm{B}}} + 1}}.
\end{equation}
When ${K_{\rm{B}}} \to \infty $, $\frac{{K_{\rm{B}}^2}}{{{K_{\rm{B}}} + 1}} \to \infty $ holds and implies that ${\bf{g}}_{{\rm{B,N}}}^H{{\bf{w}}_{{\rm{B,N}},i_{{\rm{B,L}}}^\# }}$ is negligible compared with $\sqrt {{K_{\rm{B}}}} g_{{\rm{B,1}}}^H{e^{j{\varepsilon _{i_{{\rm{B,L}}}^\# }}}}$. In this case, \eqref{eq:APP-E-2} can be approximated as
\begin{equation}\label{eq:APP-E-7}
  \begin{aligned}
{\mathbb E}\left\{ {\left| {{\bf{g}}_{\rm{B}}^H{{\bf{w}}_{{\rm{B}},i_{\rm{B}}^\# }}} \right|\left| {g_{{\rm{B,1}}}^H} \right|} \right\} \approx \sqrt {\frac{1}{{{K_{\rm{B}}} + 1}}} {\mathbb E}\left\{ {| {\sqrt {{K_{\rm{B}}}} g_{{\rm{B,1}}}^H{e^{j{\varepsilon _{i_{{\rm{B,L}}}^\# }}}}} |\left| {g_{{\rm{B,1}}}^H} \right|} \right\} = \sqrt {\frac{{{K_{\rm{B}}}}}{{{K_{\rm{B}}} + 1}}} {K_{\rm{B}}}\left( {{L_{\rm{B}}} - 1} \right).
\end{aligned}
\end{equation}
In the same way, we have
\begin{equation}\label{eq:APP-E-8}
  {\mathbb E}\left\{ {\left| {{\bf{g}}_{\rm{M}}^H{{\bf{w}}_{{\rm{M}},i_{\rm{M}}^\# }}} \right|\left| {g_{{\rm{M,1}}}^H} \right|} \right\} \approx \sqrt {\frac{{{K_{\rm{M}}}}}{{{K_{\rm{M}}} + 1}}} {K_{\rm{M}}}\left( {{L_{\rm{M}}} - 1} \right).
\end{equation}
Substituting \eqref{eq:APP-E-7} and \eqref{eq:APP-E-8} into \eqref{eq:APP-E-1} completes the proof.

\section{}\label{App:G}
To determine the range of $\varpi $, it is necessary to study the phase of $X_1$ and $X_2$. According to the definition ${X_1} = \frac{1}{{{N_{\rm{R}}}}}{\bf{g}}_{\rm{B}}^H{{\bf{w}}_{{\rm{B}},i_{\rm{B}}^\# }}{\bf{g}}_{\rm{M}}^H{{\bf{w}}_{{\rm{M}},i_{\rm{M}}^\# }}$, the phase of $X_1$ is $
  \angle {X_1} = \angle {\bf{g}}_{\rm{B}}^H{{\bf{w}}_{{\rm{B}},i_{\rm{B}}^\# }} + \angle {\bf{g}}_{\rm{M}}^H{{\bf{w}}_{{\rm{M}},i_{\rm{M}}^\# }}$.
Taking the BS-MS channel as example, we unfold ${\bf g}_{\rm B}^H$ and ${{\bf{w}}_{{\rm{B}},i_{\rm{B}}^\# }}$ to get
\begin{equation}\label{eq:APP-G-2}
  \begin{aligned}
{\bf{g}}_{\rm{B}}^H{{\bf{w}}_{{\rm{B}},i_{\rm{B}}^\# }} = \underbrace {\sqrt {\frac{{{K_{\rm{B}}}}}{{{K_{\rm{B}}} + 1}}} g_{{\rm{B}},1}^H{w_{{\rm{B,L}},i_{{\rm{B,L}}}^\# }}}_{{V_1}} + \underbrace {\sqrt {\frac{1}{{{K_{\rm{B}}} + 1}}} {\bf{g}}_{{\rm{B,N}}}^H{{\bf{w}}_{{\rm{B,N}},i_{{\rm{B,N}}}^\# }}}_{{V_2}}
\end{aligned}.
\end{equation}
Without loss of generality, we assume $\angle {V_1} \geq \angle {V_2}$. $\angle {\bf{g}}_{\rm{B}}^H{{\bf{w}}_{{\rm{B}},i_{\rm{B}}^\# }}$ lies between $\angle {V_2}$ and $\angle {V_1}$, that is,
\begin{equation}\label{eq:APP-G-3}
  \angle {V_2} \leq \angle {\bf{g}}_{\rm{B}}^H{{\bf{w}}_{{\rm{B}},i_{\rm{B}}^\# }} \leq \angle {V_1} < \angle {V_1} + \frac{\pi }{{{2^{{b_{{\rm{B,L}}}}}}}}.
\end{equation}
Furthermore, according to the quantization error in \eqref{eq:phase_error}, we have
\begin{equation}\label{eq:APP-G-4}
  \angle {V_2} > \angle {V_1} - \frac{\pi }{{{2^{{b_{{\rm{B,L}}}}}}}}.
\end{equation}
Recalling that $\angle V_1=\angle g_{{\rm{B}},1}^H{w_{{\rm{B,L}},i_{{\rm{B,L}}}^\# }}$, we merge \eqref{eq:APP-G-3} and \eqref{eq:APP-G-4} to obtain
\begin{equation}\label{eq:APP-G-5}
  \left| {\angle {\bf{g}}_{\rm{B}}^H{{\bf{w}}_{{\rm{B}},i_{\rm{B}}^\# }} - \angle {g_{{\rm{B}},1}^H{w_{{\rm{B,L}},i_{{\rm{B,L}}}^\# }}}} \right| < \frac{\pi }{{{2^{{b_{{\rm{B,L}}}}}}}}.
\end{equation}
A similar result can be derived for the RIS--MS channel, that is,
\begin{equation}\label{eq:APP-G-6}
  \left| {\angle {\bf{g}}_{\rm{M}}^H{{\bf{w}}_{{\rm{M}},i_{\rm{M}}^\# }} - \angle {g_{{\rm{M}},1}^H{w_{{\rm{M,L}},i_{{\rm{M,L}}}^\# }}}} \right| < \frac{\pi }{{{2^{{b_{{\rm{M,L}}}}}}}}.
\end{equation}
Combining \eqref{eq:APP-G-5} and \eqref{eq:APP-G-6}, we get
\begin{equation}\label{eq:APP-G-7}
  \left| {\angle {X_1} - \left( {\angle g_{{\rm{B}},1}^H{w_{{\rm{B,L}},i_{{\rm{B,L}}}^\# }} + \angle g_{{\rm{M}},1}^H{w_{{\rm{M,L}},i_{{\rm{M,L}}}^\# }}} \right)} \right| < \frac{\pi }{{{2^{{b_{{\rm{B,L}}}}}}}} + \frac{\pi }{{{2^{{b_{{\rm{B,L}}}}}}}}.
\end{equation}
According to the definition $X_2 = {{{\left( {{{\bf{g}}_{\rm{B}}} \otimes {{\bf{g}}_{\rm{M}}}} \right)}^H}{{\bf{O}}_{1}}\left( {{{\bf{w}}_{{\rm{B}},i_{\rm{B}}^\# }} \otimes {{\bf{w}}_{{\rm{M}},i_{\rm{M}}^\# }}} \right)}$, we have
\begin{equation}\label{eq:APP-G-8}
  \angle {X_2} = \angle g_{{\rm{B,1}}}^H{w_{{\rm{B,L}},i_{{\rm{B,L}}}^\# }} + \angle g_{{\rm{M,1}}}^H{w_{{\rm{M,L}},i_{{\rm{M,L}}}^\# }}.
\end{equation}
Considering that $\varpi  = \pi  - \left| {\angle {X_1} - \angle {X_2}} \right|$, we substitute \eqref{eq:APP-G-8} into \eqref{eq:APP-G-7} to finish the proof of
\begin{equation}\label{eq:APP-G-9}
  \pi  - \left( {\frac{\pi }{{{2^{{b_{{\rm{B,L}}}}}}}} + \frac{\pi }{{{2^{{b_{{\rm{M,L}}}}}}}}} \right) < \varpi  \leq \pi.
\end{equation}
Finally, the monotonicity of $\cos \varpi$ in $[0,\pi]$ proves that
\begin{equation}\label{eq:APP-G-10}
  {\mathbb E}\left\{-\cos \varpi \right\} > {\mathbb E} \left\{{\cos \left( {\frac{\pi }{{{2^{{b_{{\rm{B,L}}}}}}}} + \frac{\pi }{{{2^{{b_{{\rm{M,L}}}}}}}}} \right)}\right\}={\cos \left( {\frac{\pi }{{{2^{{b_{{\rm{B,L}}}}}}}} + \frac{\pi }{{{2^{{b_{{\rm{M,L}}}}}}}}} \right)}.
\end{equation}

\section{}\label{App:F}
Using the independence of path gains in the BS--RIS and RIS--MS channel, we get
\begin{equation}\label{eq:APP-F-1}
  \begin{aligned}
{\mathbb E}\left\{ {{{\left| {{X_1}} \right|}^2}} \right\} = \frac{1}{{N_{\rm{R}}^2}}{\mathbb E}\left\{ {{{\left| {{\bf{g}}_{\rm{B}}^H{{\bf{w}}_{{\rm{B}},i_{\rm{B}}^\# }}} \right|}^2}} \right\}{\mathbb E}\left\{ {{{\left| {{\bf{g}}_{\rm{M}}^H{{\bf{w}}_{{\rm{M}},i_{\rm{M}}^\# }}} \right|}^2}} \right\}
\end{aligned}.
\end{equation}
By decoupling the LoS and NLoS component of ${{\bf{g}}_{\rm{B}}}$ and ${{\bf{w}}_{{\rm{B}},i_{\rm{B}}^\# }}$, we get
\begin{equation}\label{eq:APP-F-2}
  \begin{aligned}
&{\mathbb E}\{ {{{| {{\bf{g}}_{\rm{B}}^H{{\bf{w}}_{{\rm{B}},i_{\rm{B}}^\# }}} |}^2}} \}  = {\mathbb E}\left\{ \frac{1}{{K_{\rm{B}}} + 1}( {\sqrt {{{{K_{\rm{B}}}}}} g_{{\rm{B,1}}}^H{e^{j{\varepsilon _{i_{{\rm{B,L}}}^\# }}}} +  {\bf{g}}_{{\rm{B,N}}}^H{{\bf{w}}_{{\rm{B,N}},i_{{\rm{B,L}}}^\# }}} )   ({\sqrt {{{{K_{\rm{B}}}}}} {g_{{\rm{B,1}}}}{e^{ - j{\varepsilon _{i_{{\rm{B,L}}}^\# }}}} + {\bf{w}}_{{\rm{B,N}},i_{{\rm{B,L}}}^\# }^H{{\bf{g}}_{{\rm{B,N}}}}})\right\}\\
& = \frac{1}{{{K_{\rm{B}}} + 1}}\left({{K_{\rm{B}}}}{\mathbb E}\left\{ {{{\left| {{g_{{\rm{B,1}}}}} \right|}^2}} \right\} + {{2\sqrt {{K_{\rm{B}}}} }}\Re\left( {\mathbb E}{\left\{ {g_{{\rm{B,1}}}^H{e^{j{\varepsilon _{i_{{\rm{B,L}}}^\# }}}}{\bf{w}}_{{\rm{B,N}},i_{{\rm{B,L}}}^\# }^H{{\bf{g}}_{{\rm{B,N}}}}} \right\}} \right) + {\mathbb E}\left\{ {{{\left| {{\bf{g}}_{{\rm{B,N}}}^H{{\bf{w}}_{{\rm{B,N}},i_{{\rm{B,L}}}^\# }}} \right|}^2}} \right\}\right).
\end{aligned}
\end{equation}
Considering that NLoS path gains have zero mean and are independent of the LoS path gain, the second term in \eqref{eq:APP-F-2} can be derived as
\begin{equation}\label{eq:APP-F-3}
  {\mathbb E}\left\{ {g_{{\rm{B,1}}}^H{e^{j{\varepsilon _{i_{{\rm{B,L}}}^\# }}}}{\bf{w}}_{{\rm{B,N}},i_{{\rm{B,L}}}^\# }^H{{\bf{g}}_{{\rm{B,N}}}}} \right\} = 0.
\end{equation}
It can be concluded that the effect of quantization bits for LoS paths vanishes here; thus, ${\mathbb E}\{ {{{| {{X_1}} |}^2}} \}$ exclusively depends on the quantization error brought by using the RVQ codebook. Utilizing formula decomposition \eqref{eq:APP-E-4} and the characteristic of RVQ codebook \eqref{eq:APP-E-5}, we can obtain
\begin{equation}\label{eq:APP-F-4}
  {\mathbb E}\left\{ {{{\left| {{\bf{g}}_{{\rm{B,N}}}^H{{\bf{w}}_{{\rm{B,N}},i_{{\rm{B,L}}}^\# }}} \right|}^2}} \right\} > \left( {{L_{\rm{B}}} - 1} \right)\left( {{K_{\rm{B}}} + 1} \right)\left( {1 - {2^{ - \frac{{{b_{{\rm{B,N}}}}}}{{{L_{\rm{B}}} - 2}}}}} \right).
\end{equation}
At this point, substituting ${\mathbb E}\left\{ {{{\left| {{g_{{\rm{B,1}}}}} \right|}^2}} \right\} = {K_{\rm{B}}}\left( {{L_{\rm{B}}} - 1} \right)$, \eqref{eq:APP-F-3}, and \eqref{eq:APP-F-4} into \eqref{eq:APP-F-2}, we have
\begin{equation}\label{eq:APP-F-5}
  {\mathbb E}\left\{ {{{\left| {{\bf{g}}_{\rm{B}}^H{{\bf{w}}_{{\rm{B}},i_{\rm{B}}^\# }}} \right|}^2}} \right\} > \left( {{L_{\rm{B}}} - 1} \right)\left(\frac{{K_{\rm{B}}^2}}{{{K_{\rm{B}}} + 1}} + 1 - {2^{ - \frac{{{b_{{\rm{B,N}}}}}}{{{L_{\rm{B}}} - 2}}}} \right).
\end{equation}
Using the same derivation process, a similar result can be obtained for the RIS--MS channel, that is,
\begin{equation}\label{eq:APP-F-6}
  {\mathbb E}\left\{ {{{\left| {{\bf{g}}_{\rm{M}}^H{{\bf{w}}_{{\rm{M}},i_{\rm{M}}^\# }}} \right|}^2}} \right\} > \left( {{L_{\rm{M}}} - 1} \right)\left(\frac{{K_{\rm{M}}^2}}{{{K_{\rm{M}}} + 1}} +  {1 - {2^{ - \frac{{{b_{{\rm{M,N}}}}}}{{{L_{\rm{M}}} - 2}}}}} \right).
\end{equation}
Finally, \emph{Theorem \ref{Them:2}} can be proved by bringing \eqref{eq:APP-F-5} and \eqref{eq:APP-F-6} back to \eqref{eq:APP-F-1}.

\begin{small}

\end{small}
%\section{Unused characters}
%$d,e$\par
%$m$\par
%$o,p,q,r,t$\par
%$u,z$\par
%
%$N, Q,R, \Delta, {\bf H}, K,g,\Theta,\Phi,{\bf{a,f}}$


\begin{thebibliography}{30}

\bibitem{smart-radio}M. Di Renzo \emph{et al.}, ``Smart radio environments empowered by reconfigurable AI meta-surfaces: An idea whose time has come,'' \emph{EURASIP J. Wireless Commun. Networking}, 2019. vol. 2019, no. 129, May 2019.

\bibitem{Y. Han-LIS}Y. Han \emph{et al.}, ``Large intelligent surface-assisted wireless communication exploiting statistical CSI,'' \emph{IEEE Trans. Vehicular Tech.,} vol. 68, no. 8, pp. 8238-8242, Aug. 2019.

\bibitem{Q. Wu-IRS}Q. Wu and R. Zhang, ``Intelligent reflecting surface enhanced wireless network via joint active and passive beamforming,'' \emph{IEEE Tans. Wireless Commun.}, Vol. 18, no. 11, pp. 5394-5409, Nov. 2019.

\bibitem{S. Gong}S. Gong \emph{et al.}, ``Towards smart radio environment for wirelsee communications via intelligent reflecting surfaces: A comprehensive sruvey,'' [Online].  Available: https://arxiv.org/abs/1912.07794v1


\bibitem{W. Tnag_1}W. Tang, J. Y. Dai, M. Chen, X. Li, Q. Cheng, S. Jin, K.-K. Wong, and T. J. Cui, ``Programmable metasurface-based RF chain-free 8PSK wireless transmitter,'' \emph{IET Electron. Lett.,} vol. 55, no. 7, pp. 417-420, Apr. 2019.
\bibitem{W. Tnag_2}W. Tang, X. Li, J. Y. Dai, S. Jin, Y. Zeng, Q. Cheng, and T. J. Cui, ``Wireless communications with programmable metasurface: Transceiver design and experimental results,'' \emph{China Commun.,} vol. 16, no. 5, pp. 46-61, May 2019.
\bibitem{W. Tnag_3}W. Tang \emph{et al.}, ``Wireless communications with programmable metasurface: New paradigms, opportunities, and challenges on transceiver design,'' \emph{IEEE Wireless Commun Mag.}, vol. 27, no. 2, pp. 180-187, Apr. 2020.
\bibitem{W. Tnag_4}W. Tang \emph{et al.}, ``MIMO transmission through reconfigurable intelligent surface: System design, analysis, and implementation,'' \emph{IEEE J. Sel. Commun.}, 2020. [Online]. Available: https://arxiv.org/abs/1912.09955.

\bibitem{C. Liaskos}C. Liaskos, S. Nie, A. Tsioliaridou, A. Pitsillides, S. Ioannidis, and I. Akyildiz, ``A new wireless communication paradigm through software-controlled metasurfaces," \emph{IEEE Commun. Mag.,} vol. 56, no. 9, pp. 162-169, Sept. 2018.

\bibitem{App_EE}C. Huang \emph{et al.}, ``Reconfigurable intelligent surfaces for energy efficiency in wireless communication,'' \emph{IEEE Trans. Wireless Commun.,} vol. 18, no. 8, pp. 4157-4170, Aug. 2019.

%\bibitem{App_localization}A. Elzanaty and A. Guerra, ``Reconfigurable Intellegent surfaces for localization: Position and orientation error bounds,'' [Online]. Available: https://arxiv.org/abs/2009.02818v1.

\bibitem{App_scerecy}F. Shu \emph{et al.}, ``Enhanced secrecy rate maximization for directional modulation networks via IRS,'' [Online]. Available: https://arxiv.org/abs/2008.05067v1.

%\bibitem{App_NOMA}L. Bariah \emph{et al.}, ``Large intelligent surface assisted non-orthogonal multiple access: Performance analysis,'' [Online]. Available: https://arxiv.org/abs/2007.09611v.

\bibitem{App-reflecting-modulation}S. Guo, S. Lv, H. Zhang, J. Ye, and P. Zhang, ``Refleting Modulation,'' [Online]. Available: https://arxiv.org/abs/1912.08428v1.

\bibitem{App-mmWave-1}M. Nemati, J. Park, and J. Choi, ``RIS-assisted coverage enhancement on millimeter-wave cellular networks,'' [Online]. Available: https://arxiv.org/abs/2007.08196v1.

\bibitem{App-mmWave-2}W. Chen, X. Yang, S. Jin, and P. Xu, ``Sparse array of sub-surface aided anti-blockage mmWave communication systems,'', \emph{in Proc. IEEE Globecom2020}, 2020.

\bibitem{App-D2D}Y. Chen \emph{et al.}, ``Reconfigurable intelligent surface assisted device-to-device communications,'' [Online]. Available: https://arxiv.org/abs/2007.00859v1.

\bibitem{App-IoT}Q. Zhang, Y.-C. Liang, and H. V. Poor, ``Large intelligent surface/antennas (LISA) assisted symbiotic radio for IoT communications,'' [Online]. Available: https://arxiv.org/abs/2002.00340v1.

\bibitem{est-atomic}J. He, H. Wymeersch, and M. Juntti, ``Channel estimation for RIS-aided mmWave MIMO systems via atomic norm minimization,'' [Online]. Available: https://arxiv.org/abs/2007.08158v1.

\bibitem{est-THz}B. Ning, Z. Chen, W. Chen, and Y. Du, ``Channel estimation and transmission for intelligent reflecting surface assisted THz communications,''\emph{ in Proc. 2020 IEEE International Conference on Communications (ICC),} Dublin, Ireland, 2020.

\bibitem{est-partial}B. Zheng, C. You, and R. Zhang, ``Intelligent reflecting surface assisted multi-user OFDMA: Channel estimation and training design,'' [Online]. Available: https://arxiv.org/abs/2003.00648v1.

\bibitem{est-CS}J. Chen, Y.-C. Liang, H. V. Cheng, and W. Yu, ``Channel estimation for reconfigurable intelligent surface aided multi-user MIMO systems,'' [Online]. Available: https://arxiv.org/abs/1912.03619v1.

\bibitem{est-sensor}A. Taha, M. Alrabeiah, and A. Alkhateeb, ``Enabling large intelligent surfaces with compressive sensing and deep learning,'' [Online]. Available: https://arxiv.org/abs/1904.10136v2.

\bibitem{Nihar}N. Jindal, ``MIMO broadcast channels with finite-rate feedback,'' \emph{IEEE Trans. Inf. Theory}, vol. 52, no. 11, pp. 5045-5060, Nov. 2006.

\bibitem{Ramta_1}R. Bhagavatula and R. W. Heath, ``Adaptive limited feedback for sum-rate maximizing beamforming in cooperative multicell systems,'' \emph{IEEE Trans. signal Process.}, vol. 59, no.2, pp. 800-811, Feb. 2011.

\bibitem{Ramta_2}R. Bhagavatula and R. W. Heath, ``Adaptive bit partitioning for multicell intercell interference nulling with delayed limited feedback,'' \emph{IEEE Trans. signal Process.}, vol. 59, no.8, pp. 3824-3836, Aug. 2011.

\bibitem{feedback-RIS}D. Shen and L. Dai, ``Channel feedback for reconfigurable intelligent surface assisted wireless communications,'' [Online]. Available: https://arxiv.org/abs/2004.07174v1.

\bibitem{Naive_RVQ}W. Shen, L. Dai, B. Shim, Z. Wang, and R. W. Heath, Jr., ``Channel feedback based on AoD-adaptive subspace codebook in FDD massive MIMO systems,'' \emph{IEEE Trans. Commun.,} vol. 66, no. 11, pp. 5235-5248, Nov. 2018.

\bibitem{Han_FDD}Y. Han, T. Hsu, C. Wen, K. Wong, and S. Jin, ``Efficient downlink channel reconstruction for FDD multi-antenna systems,'' \emph{IEEE Trans. Wireless Commun.}, vol. 18, no. 6, pp. 3161-3176, June 2019.

\bibitem{SAoS}W. Chen, X. Yang, S. Jin and P. Xu, ``Sparse array of sub-surface aided block-free multi-user mmWave communication systems,'' \emph{Digit. Commun. Netw.,} vol. 6, no. 3, pp. 292-303, Aug. 2020.

\bibitem{AngleReci}W. Chen, L. Bai, W. Tang, S. Jin, W. X. Jiang, and T. J. Cui, ``Angle-dependent phase shifter model for reconfigurable intelligent surfaces: Does the angle-reciprocity hold?'' \emph{IEEE Commun. Lett.}, vol. 24, no. 9, pp. 2060-2064, Sept. 2020.

\bibitem{beam_positioning}W. Chen, S. He, Q. Xu, J. Ren, Y. Huang and L. Yang, ``Positioning algorithm and AoD estimation for mmWave FD-MISO system,'' in \textit{Proc. 2018 IEEE International Conference on Wireless Communications and Signal Processing,} 2018, pp. 1-6.


\end{thebibliography}
\end{document}